\documentclass{aa}
\usepackage{epsfig}
\usepackage{longtable} 
\usepackage{graphics}

\newbox\grsign \setbox\grsign=\hbox{$>$} \newdimen\grdimen \grdimen=\ht\grsign
\newbox\simlessbox \newbox\simgreatbox \newbox\simpropbox \newbox\wtildebox 
\setbox\simgreatbox=\hbox{\raise.5ex\hbox{$>$}\llap
     {\lower.5ex\hbox{$\sim$}}}\ht1=\grdimen\dp1=0pt
\setbox\simlessbox=\hbox{\raise.5ex\hbox{$<$}\llap
     {\lower.5ex\hbox{$\sim$}}}\ht2=\grdimen\dp2=0pt

\def\simless{\mathrel{\copy\simlessbox}}

\newcommand{\kms}{\mbox{km~s$^{-1}$}}

\newcommand{\vedge}{\mbox{$v_{\rm edge}$}}
\newcommand{\Rsun}{\mbox{$R_{\odot}$}}

\begin{document} 

\title{A study of the expanding envelope of Nova V1974 Cyg 1992 based
on IUE high resolution spectroscopy}

\author{
Angelo Cassatella\inst{1,2}
\and Henny J.G.L.M. Lamers \inst{3} 
\and Corinne Rossi\inst{4}
\and Aldo Altamore \inst{2}
\and Rosario Gonz\'alez--Riestra\inst{5}
}
\offprints{A. Cassatella}
\mail{cassatella@fis.uniroma3.it}

\institute { Istituto di Astrofisica Spaziale e Fisica Cosmica, 
CNR, Area di Ricerca
Tor Vergata, Via del Fosso del Cavaliere 100, 00133 Roma, Italy 
\and
Dipartimento di Fisica E. Amaldi, Universit\`a degli Studi Roma Tre,
Via della Vasca Navale 84, 00146 Roma, Italy 
\and Astronomical Institute and SRON Laboratory for Space Research, 
Utrecht University, Princetonplein 5, NL3584CC, Utrecht, The Netherlands
\and Dipartimento di
Fisica, Universit\`a degli Studi La Sapienza, Piazzale A. Moro
2, 00185 Roma, Italy  
\and XMM Observatory, Villafranca Satellite Tracking Station,
 P.O.  Box 50727, 28080 Madrid, Spain
}

\date{Received / Accepted}

\authorrunning{A. Cassatella et al.}
\titlerunning{Nova V1974 Cyg: IUE high resolution spectroscopy}

%**********************************************************************
\abstract{We have carried out a detailed analysis of the ${\it IUE}$
archival high resolution spectra of the classical nova V1974 Cyg 1992.
The main UV resonance lines show P Cygni profiles in the first days,
which change into symmetric pure emission lines, and then slowly
become fainter and narrower. Lines of higher ionization species reach
their peak luminosity later than those of low ionization.  This can be
explained by a fast wind which is optically thick in the early days,
when the pseudo-photosphere  is located inside the wind. As the mass loss
decreases, the radius of the pseudo-photosphere schrinks. This has
three effects that explain the observed changes: (1) the deeper
accelerating layers of the wind become visible where the emission
lines are formed by collisional excitation and/or recombination, (2)
as the mass loss rate decreases the emission comes from deeper regions of 
the wind where the velocities are smaller, (3) the effective
temperature and the degree of ionization increase.  In addition to the
P Cygni and emission lines, we could identify two shortward shifted
absorption systems which originate in two separate expanding shells,
outside the wind layers where the emission lines are formed. The
velocity of both shells increase with time.  The outer main shell,
containing most of the matter ejected at the outburst, produces the
so-called ``principal absorption line system'', and the inner faster
moving second shell produces the so-called ``diffuse--enhanced
absorption line system''.  The acceleration of the two shells is the
result of increasing line-radiation pressure due to the UV-brightening
of the star as the effective radius decreases.  Around day 60 the
second shell has overtaken the slower moving principal system shell,
and merged with it. This explains: the sudden disappearance of the
diffuse line system near that date, the upward jump of ${\Delta}v$ =
240 km~s$^{-1}$ in velocity of the principal system and the first
detection of hard X--ray emission on day 63. This velocity jump
indicates that the main shell is ${\approx}$ 4 times more massive than
the second shell. The deceleration suffered by the diffuse--enhanced
system after the shock provides a shock temperature $T_{shock}$
$\approx$ 1.6~ keV, in fairly good agreement with the temperature of
the observed hard X--ray emission.  The UV observations are
interpreted through an empirical model in which the pre--nova slow
wind phase is followed by the ejection of two shells, where the
principal and the diffuse--enhanced absorption systems are formed, and
by a phase of fast continuous lower density wind.  Our empirical
expansion velocity law for the principal system, together with
H${\alpha}$ interferometric observations of the angular radius on day
10 are used to determine the distance to the nova, which is found to
be 2.9 $\pm$ 0.2 kpc, in agreement with {\it HST} imaging and with the
absolute magnitude versus rate of decline
relationship. 
}
\maketitle

%**********************************************************************
%**********************************************************************

\section{Introduction}
\label{sec:intro}

V1974 Cyg (Nova Cygni 1992) was discovered in outburst by Collins
(1992) on 1992 February 19.07 UT (J.D. = 2448672.3) at a visual
magnitude $V$ = 6.8.  The {\it discovery time} will be taken as day
zero all through this paper.  

The true time of the outburst of V1974 Cyg can be roughly estimated by
comparing its visual light curve, characterized by an optical decay
time $t_3$ = 42 $\pm$ 2 days (Chochol et al. 1993), with that of two
novae having a very similar value of $t_3$ of about 40 days: N Lac
1910.9 and N Mon 1918.0. The rise time from quiescence to maximum is
${\approx}$ 8 days for both objects (cf. the light curves in Cecchini
\& Gratton 1942).  We can then estimate, by analogy, that the outburst
of V1974 Cyg took place about 8 days before optical maximum,
i.e. about 4.6 days before the reported discovery.  An independent
guess of the effective date of outburst can be made by extrapolating
the pre-maximum light curve of V1974 Cyg to the magnitude at
quiescence, as quoted in Andrillat \& Houziaux (1993). We obtain in
this case that the outburst took place at least 2 days before the time
of discovery.  By averaging the two estimates, we obtain that the
outburst has probably taken place about 3.3 days before the discovery
time (our adopted day zero).

V1974 Cyg has immediately caught the attention of observers because of
its exceptional brightness at maximum ($V_{max} \approx$ +4.4 visual
magnitude) reached on February 22.5 (Rosino et al. 1996) and
consequently, as being an easy target for multiwavelength
observations, which were soon extended to the ultraviolet, radio,
infrared and X--rays. These abundant data represent a unique
opportunity to get insight into important questions concerning the
nova phenomenon.

In this paper we study the dynamics of the expanding envelope of V1974
Cyg through the analysis of the velocity variations of the ultraviolet
absorption lines.  Our main purpose is to get insight on the velocity
law of the absorption systems, on their origin and possible
interaction, and on the velocity and ionization stratification within
the envelope.

In general the optical spectra of classical novae show four systems of
shortward shifted absorption lines (McLaughlin 1960). In order of
appearance, of increasing radial velocity and degree of ionization
these systems are: the pre--maximum, the principal, the
diffuse--enhanced and the Orion systems.  (For a critical review of
the controversial interpretation of the absorption line systems in
novae see Friedjung \& Duerbeck (1993)). In this paper we will adopt
this classification scheme.

%**********************************************************************
%**********************************************************************

\section{Observations}
\label{sec:observa}

%-----------------------------------------------------------------------
\begin{figure}
\begin{center}
\psfig{file=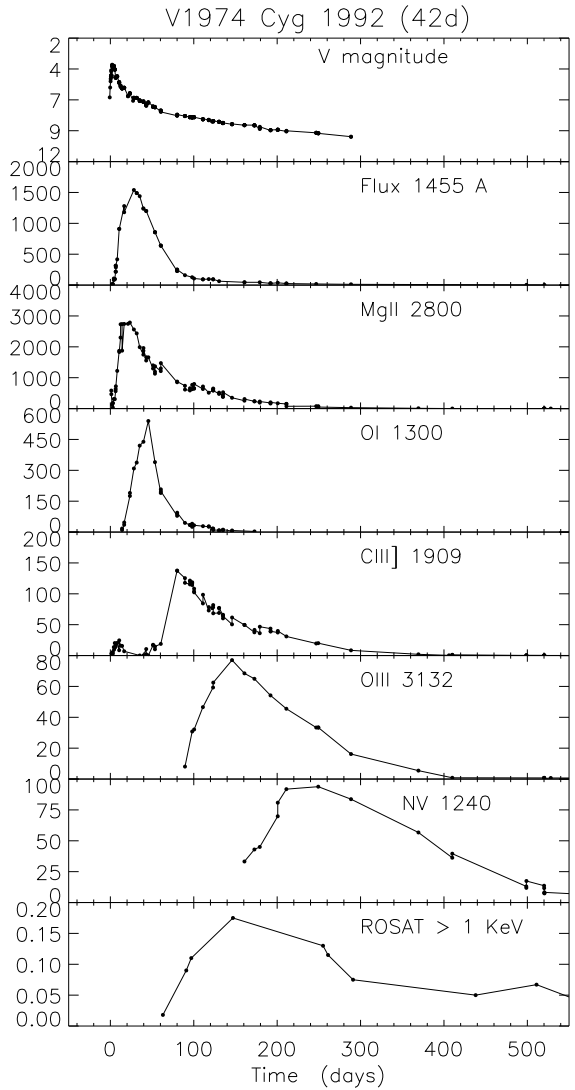,width=8cm}
\caption{Temporal evolution of the visual magnitude (from IAU
circulars), of the observed fluxes in the 1455 \AA\ continuum and of
some strong emission lines (from {\it IUE} low resolution spectra), 
and of the hard X--ray PSPC {\it ROSAT} emission (counts/s) (from
Balman et al. 1998). {\it IUE} fluxes in the continuum and in the
emission lines are in units of 10$^{-12}$ erg cm$^{-2}$ s$^{-1}$
A$^{-1}$ and of 10$^{-12}$ erg cm$^{-2}$ s$^{-1}$, respectively}
\label{fig:lowres} 
\end{center}
\end{figure}
 
The  difficulty of line identifications in the ultraviolet
spectra of novae during early outburst is sensibly alleviated in the
case of V1974 Cygni thanks to the very large amount of spectra (about
330) obtained with the {\it IUE} satellite at both high and low
resolution covering roughly 0.8 and 3.3 years of the nova evolution,
respectively.

We mainly concentrate on the analysis of the high resolution
observations of V1974 Cyg (about 140 spectra), which cover 200 days
and 330 days of its post--outburst evolution in the SWP and LWP
cameras, respectively, at a resolution of about 25 \kms.  However,
there is unfortunately an observational gap between day 53 and 60,
where important changes have probably taken place, as discussed later.
The data were retrieved from the {\it INES} ({\it IUE} Newly
Extracted Spectra) system through its Principal Centre at {\tt
http://ines.laeff.esa.es}.  A full description of the {\it INES}
system for high resolution spectra is given in Gonz\'alez--Riestra et
al. (2000).

We have also used {\it IUE} low resolution spectra to highlight
specific aspects of the flux variations which cannot be studied at
high resolution given the ${\approx}$ 90 times lower detection
efficiency.  In Fig. \ref{fig:lowres} we plot as a function of time
the ultraviolet continuum flux in a band 20 \AA\ wide centered at 1455
\AA, and the flux of some strong emission lines covering a wide range
of ionization energy. 
A characteristic feature of the emission
lines is that the higher  their degree of ionization, the later they
reach a flux maximum: the maximum emission takes place on day 21 for
\ion{Mg}{ii} 2800 \AA, day 42 for \ion{O}{i} 1300 \AA, day 77 for
\ion{C}{iii]} 1909 \AA, day 143 for \ion{O}{iii} 3132 \AA, and day 215
for \ion{N}{v} 1240 \AA.
Note that the \ion{O}{iii} 3132 \AA\ line
is one of the cascade transitions produced via the Bowen fluorescence
excitation of \ion{O}{iii} by \ion{He}{ii} Lyman ${\alpha}$ (Saraph \&
Seaton 1980), and can thus be   used to track the \ion{He}{ii} 1640 \AA\
flux variations. (The \ion{He}{ii} line itself is not easily
measurable at low resolution.)  Similarly, the OI line is pumped by
fluorescence of Lyman $\beta$ and therefore reflects the strength of
the Lyman $\beta$ emission (see Cassatella et al. 2002, hereafter Paper I).

In Fig. \ref{fig:lowres} we show also, for further reference, the
optical light curve  (top; from the IAU Circulars) and the hard
X--ray flux (bottom; from Krautter et al. 1996).  Additional
information on the UV spectral evolution of V1974 Cyg (and other
novae) from ${\it IUE}$ low resolution spectra can be found in
Paper I.

%-----------------------------------------------------------------------
%**********************************************************************

\section{Spectral evolution of the P Cygni profiles and the UV emission lines}
\label{sec:evolution}

\begin{figure*}
\begin{center} 
\psfig{file=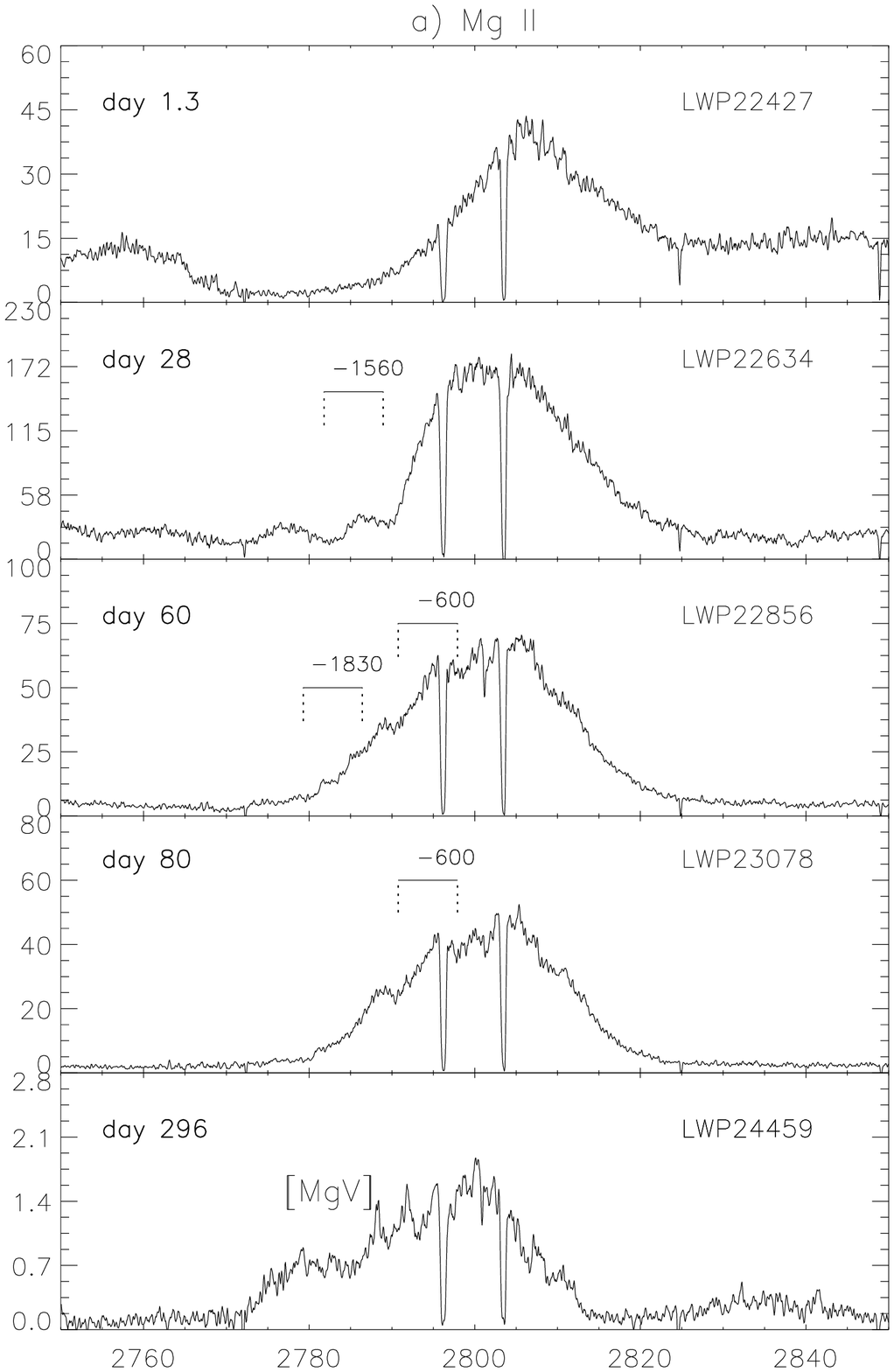,width=7.5cm}
\psfig{file=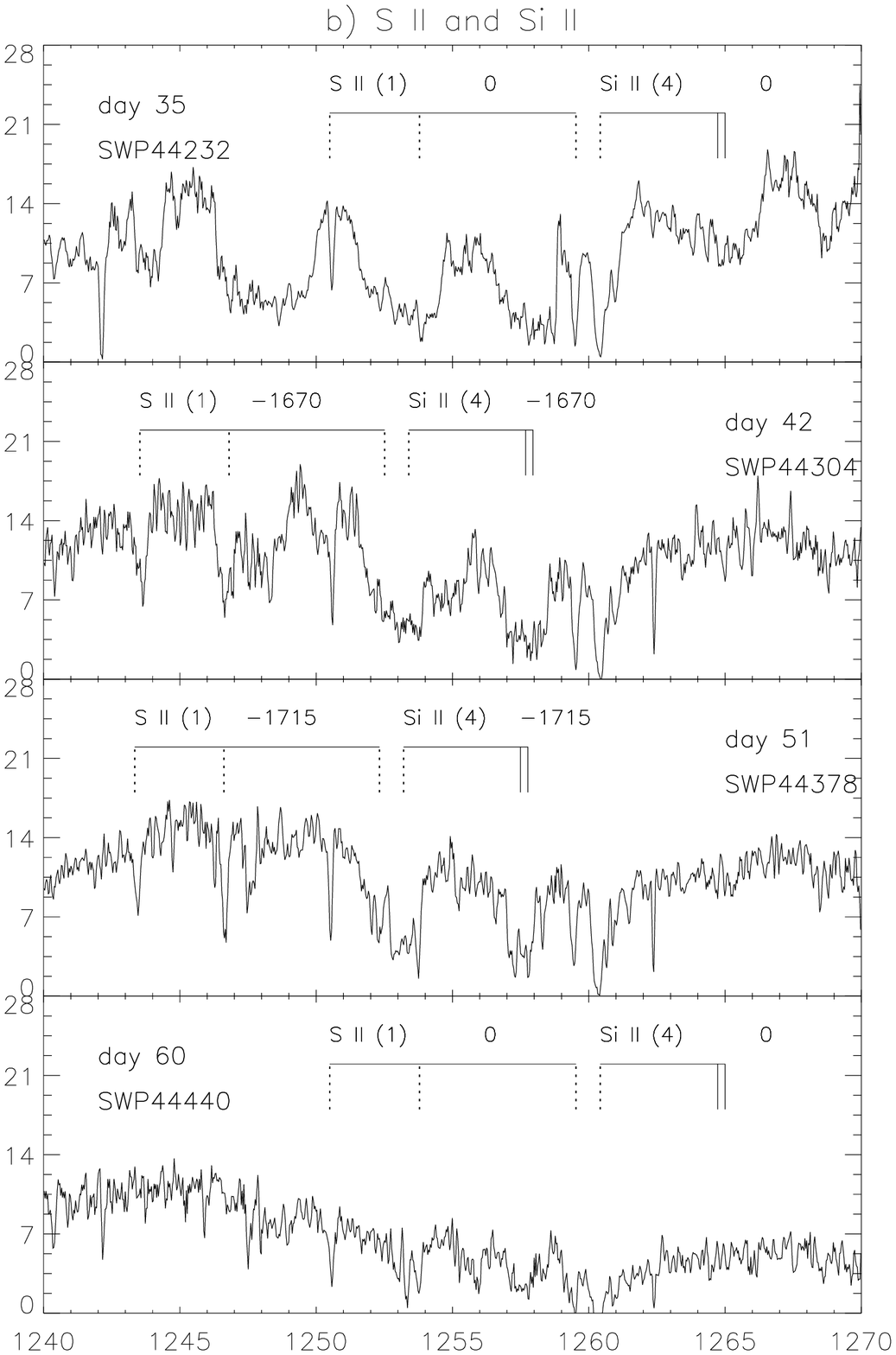,width=7.5cm}
\psfig{file=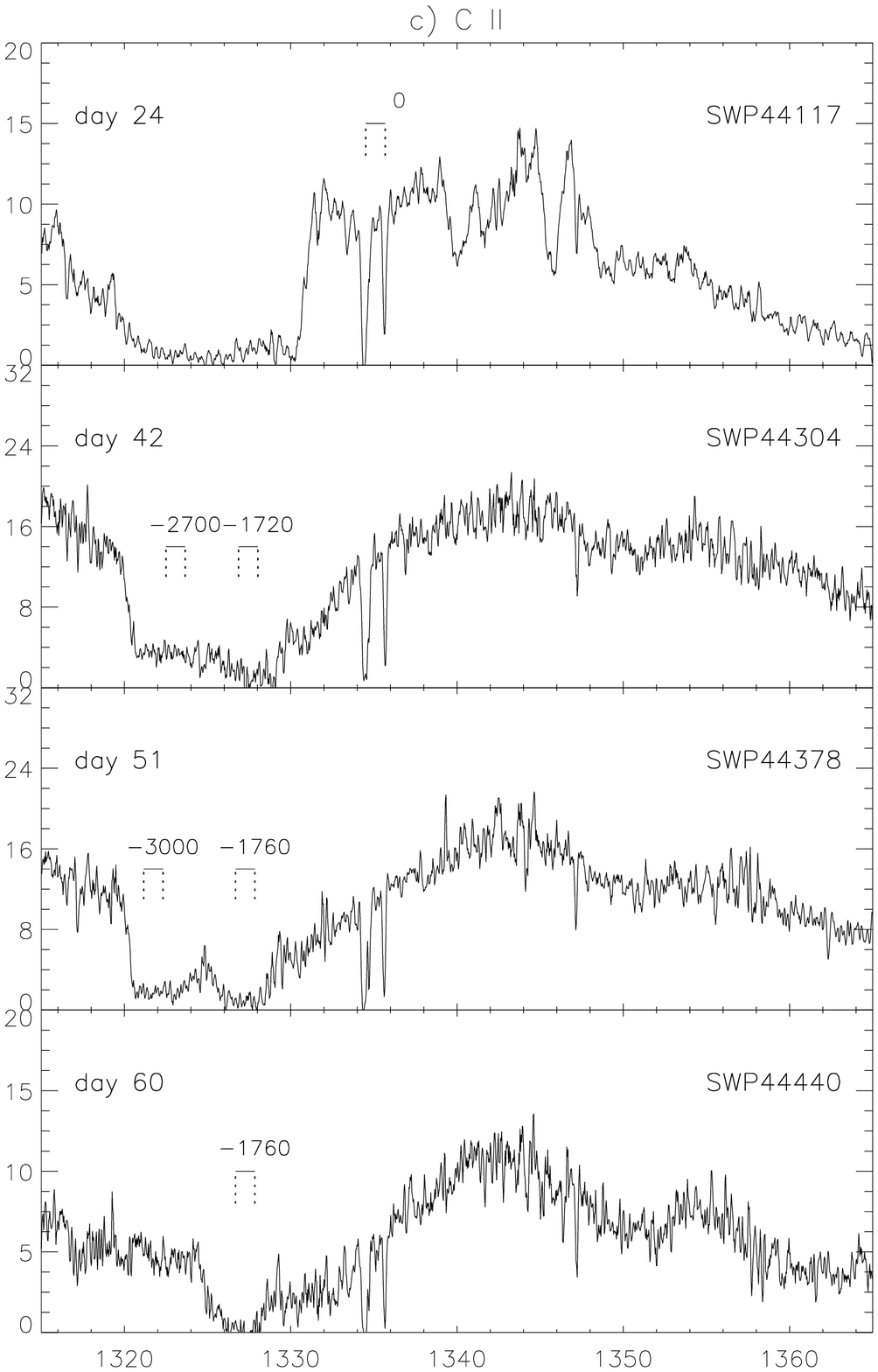,width=7.5cm}
\psfig{file=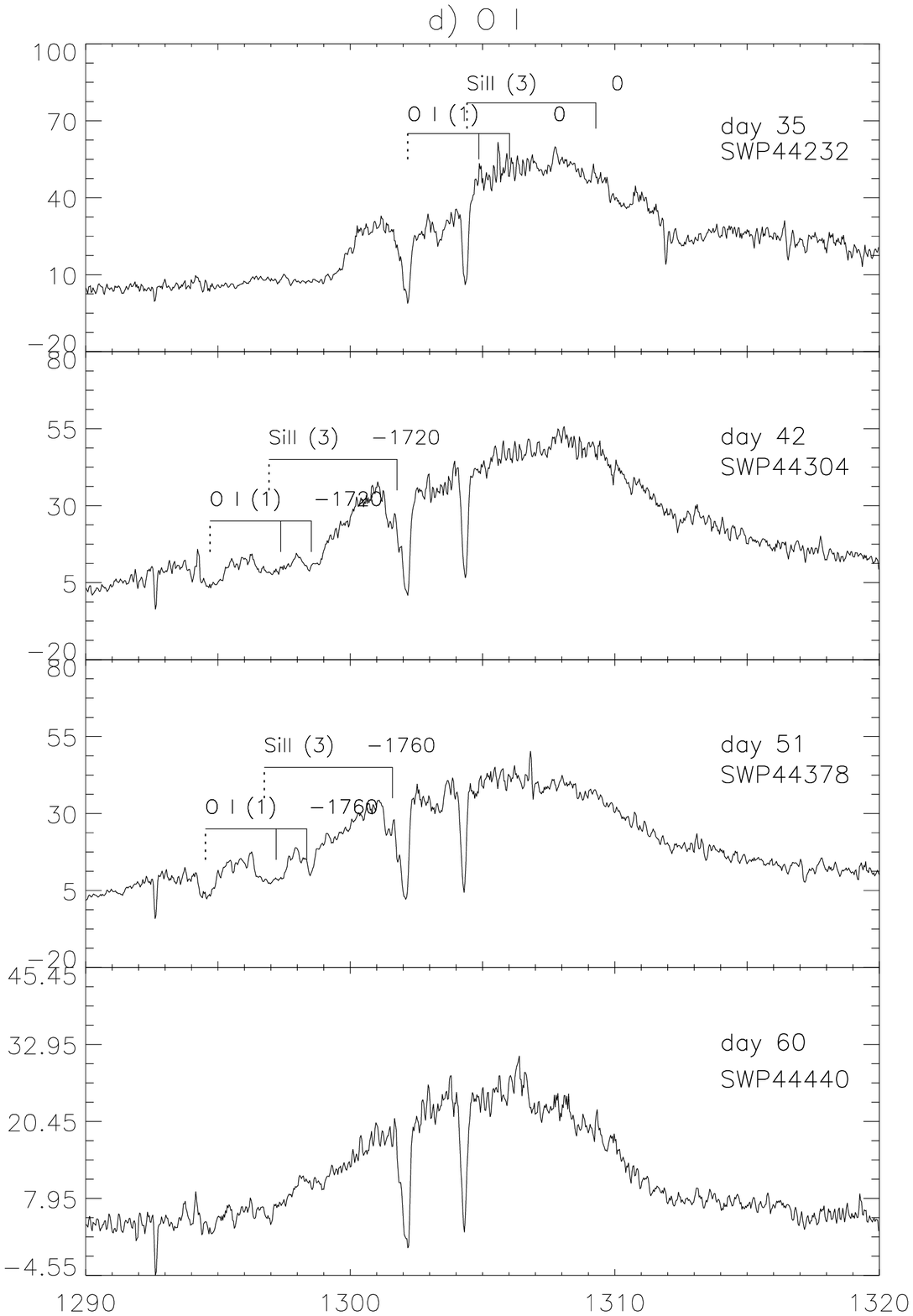,width=7.5cm}
\caption{Evolution of the line profiles in the region of \ion{Mg}{ii},
\ion{S}{ii} and \ion{Si}{ii} 1250-1265 \AA, \ion{C}{ii} 1335 \AA, and
\ion{O}{i} 1300 \AA.  The x and y axes report wavelengths in \AA\ and
fluxes in units of 10$^{-12}$ erg cm$^{-2}$ s$^{-1}$ A$^{-1}$.  The
main absorption features and the corresponding radial velocities are
indicated for the individual lines or multiplets (0 labels the
laboratory wavelength, but sometimes is omitted).  Transitions from
the 0 eV level are indicated as vertical dotted lines}
\label{fig:prof1}
\end{center} 
\end{figure*}

\begin{figure*}
\begin{center}
\psfig{file=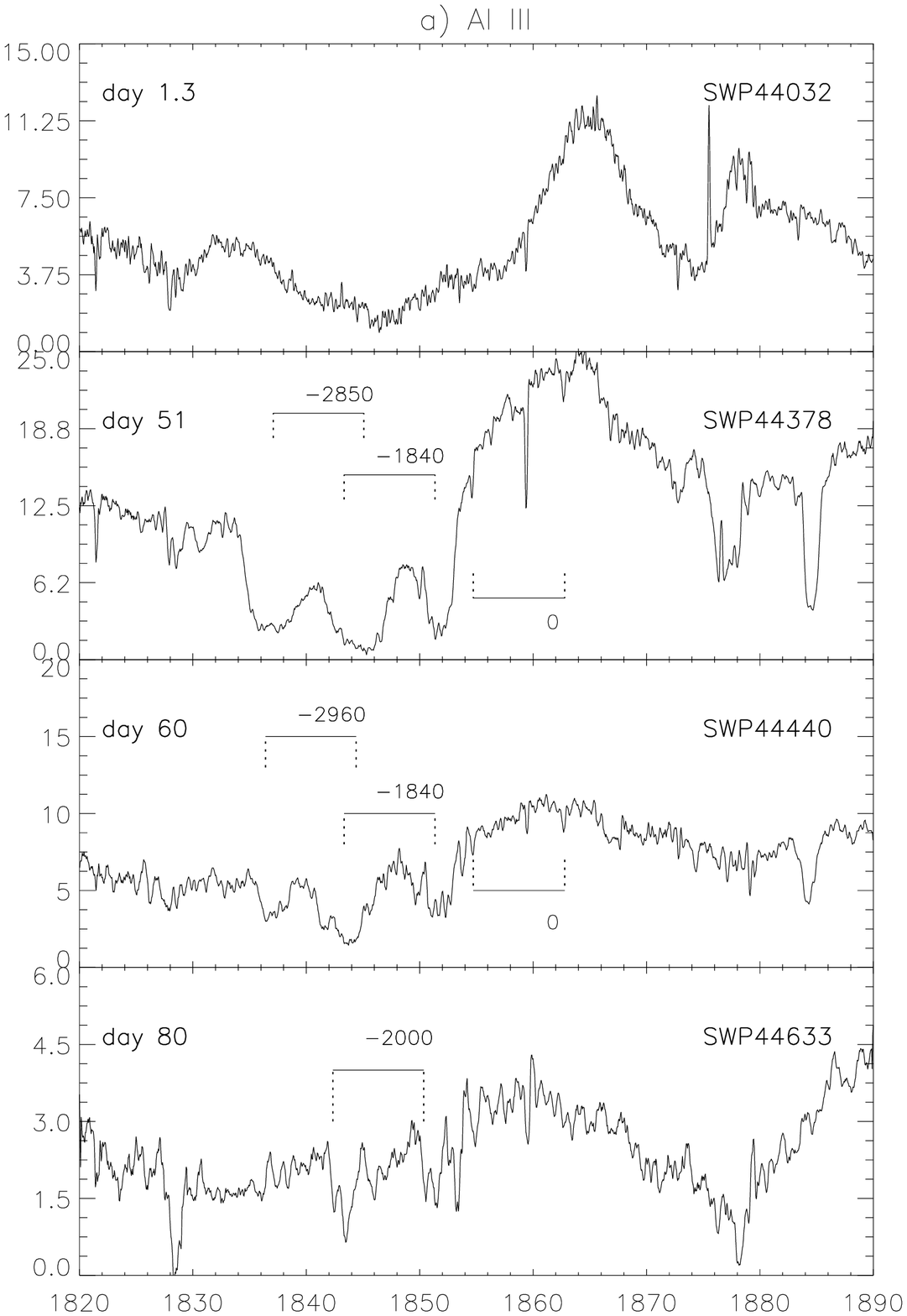,width=7.5cm}
\psfig{file=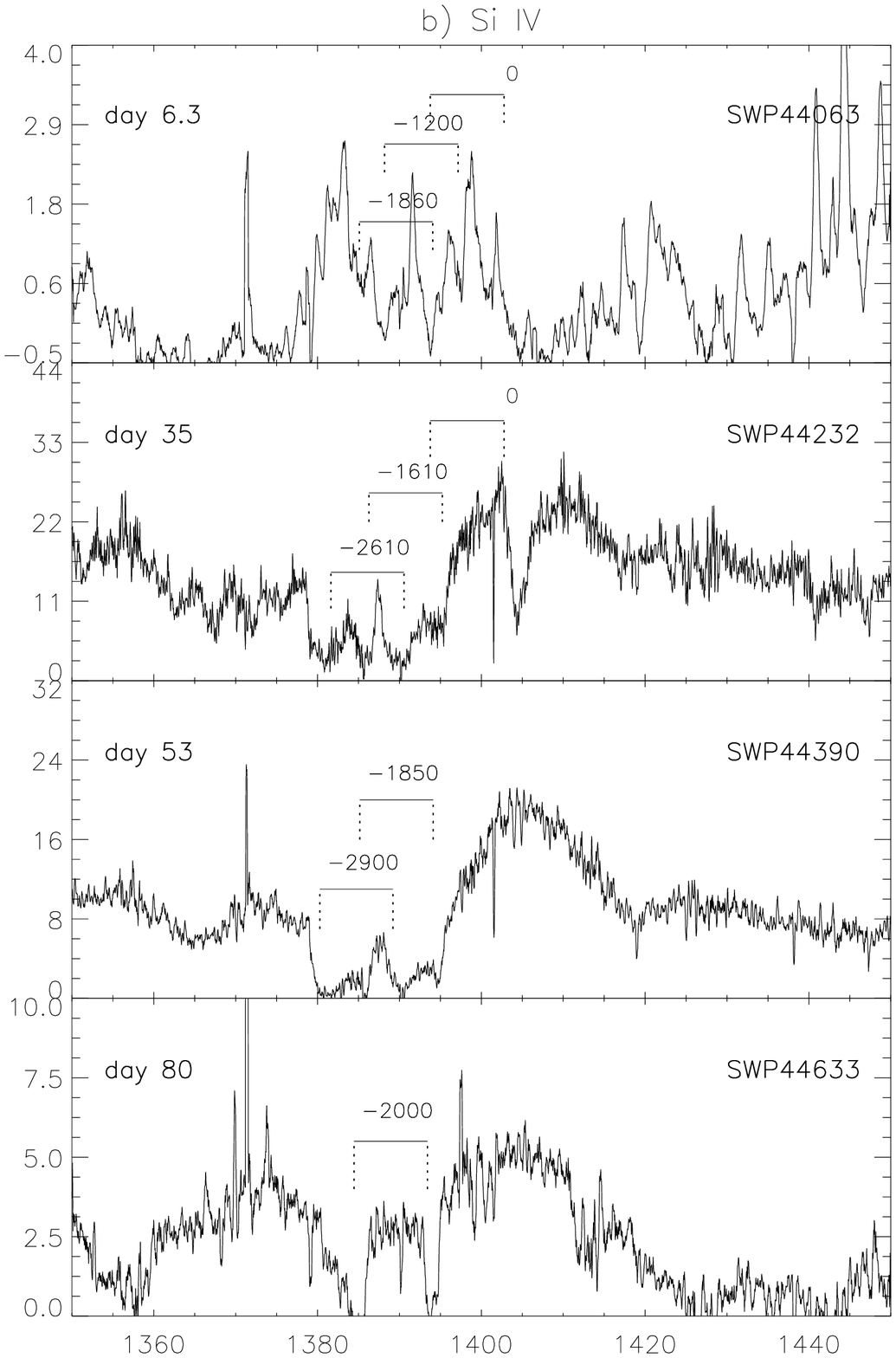,width=7.5cm}
\psfig{file=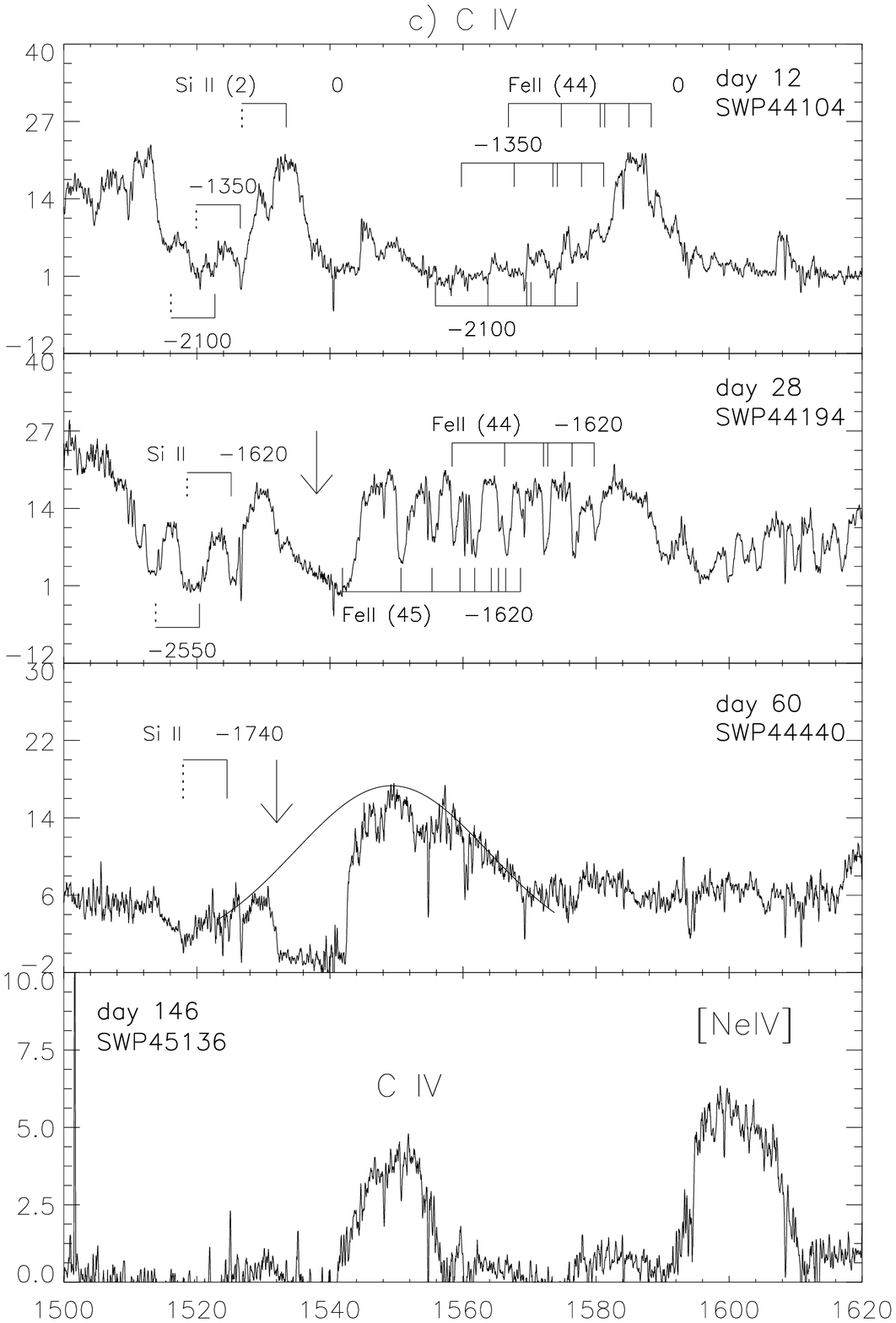,width=7.5cm}
\psfig{file=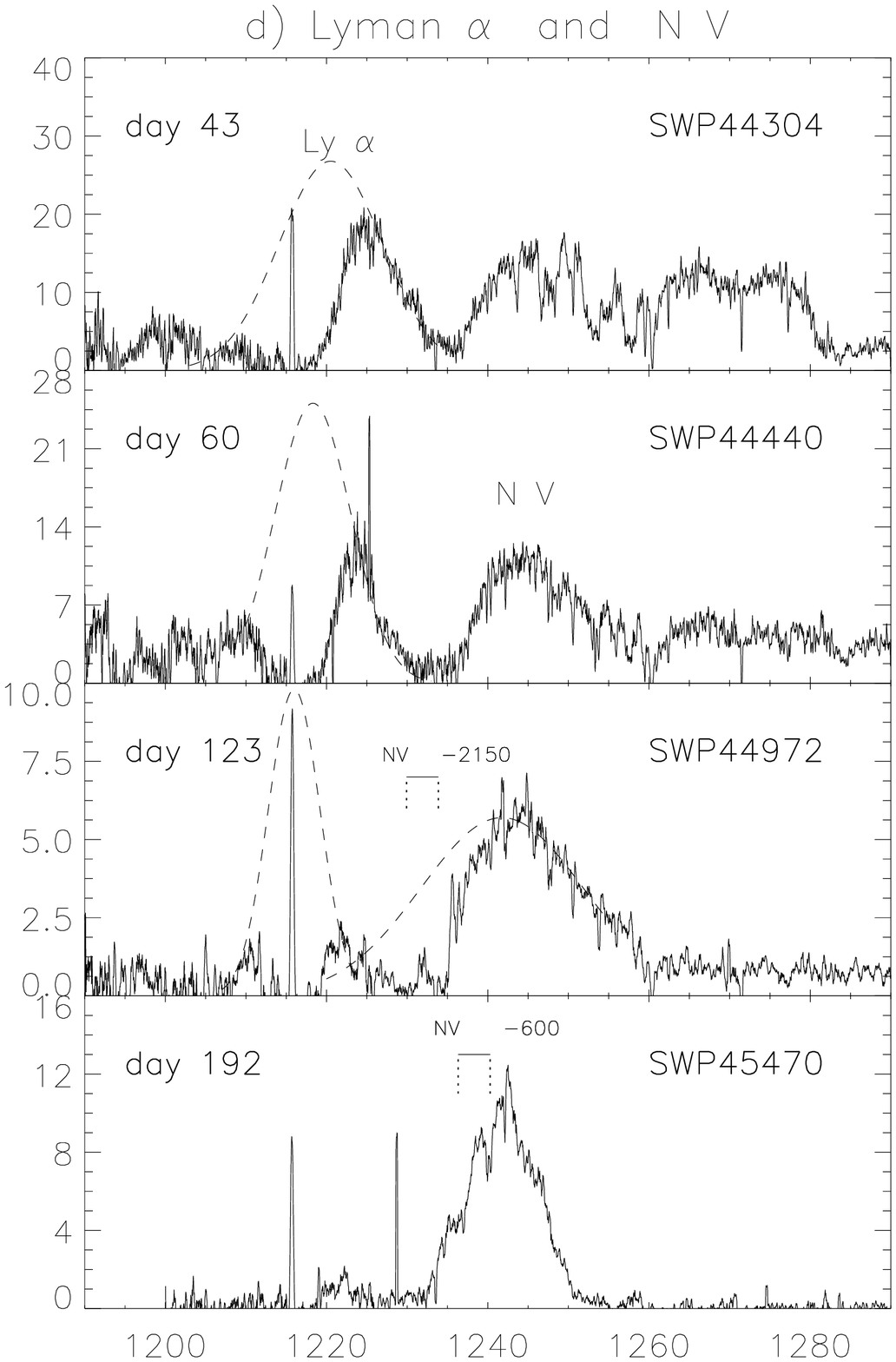,width=7.5cm}
\caption{Same as as in Fig. \ref{fig:prof1} for the lines of
\ion{Al}{iii}, \ion{Si}{iv}, \ion{C}{iv}, \ and \ion{N}{v}.  See
Fig. \ref{fig:prof4} for a complete identification of the \ion{C}{iv}
region on day 28 including the diffuse--enhanced absorption lines from
\ion{Fe}{ii}. The position of the absorption through of the
\ion{C}{iv} line on day 28 and the wavelength corresponding to its
edge velocity on day 60 are indicated with an arrow.  Panel (d) shows
the profiles of hydrogen Lyman ${\alpha}$ and \ion{N}{v} at four
epochs. To help in the identifications, we have indicated with a
dashed line a rough approximation  of these emissions with a Gaussian
function.  Note the progressive narrowing of the two emission lines,
and the presence of absorption components from the principal system in
the short wavelength wing of \ion{N}{v} on day 123, and of low
velocity components at ${\approx}$ -600 km~s$^{-1}$ on day 192 (see
Sect. \ref{sec:n1240}) }
\label{fig:prof2}
\end{center} 
\end{figure*}

In the following we discuss the spectroscopic changes of the
following lines: \ion{Mg}{ii} 2800 \AA, \ion{S}{ii} and \ion{Si}{ii}
1250-1265 \AA, \ion{C}{ii} 1335 \AA, \ion{O}{i} 1300 \AA,
\ion{Al}{iii} 1860 \AA, \ion{Si}{iv} 1400 \AA, \ion{C}{iv} 1550 \AA,
Lyman ${\alpha}$ and \ion{N}{v} 1240 \AA, which are shown in
Figs. \ref{fig:prof1} and \ref{fig:prof2}.

%-----------------------------------------------------------------------

%profilo MgII  mg2.plo  size 17x24 

%-----------------------------------------------------------------------

%profilo OI e SiII o1300.plo si1260.plo size 17x24  
%-----------------------------------------------------------------------

\subsection{P Cygni lines}
\label{sec:pcygni}

The UV spectra of V1974 Cyg generally show prominent P Cygni profiles
in the  resonance lines, especially at the earliest stages of the
evolution.  The strongest of these features  is the \ion{Mg}{ii}
2796.35, 2803.52 \AA\ doublet. It is useful to take it as a prototype
for the evolution of the other P Cygni lines because it lies in a
region of relatively low line blanketing and is thus   easier to
interpret. 

%**********************************************************************

\subsubsection{\ion{Mg}{ii} 2796.35, 2803.52 \AA}
\label{sec:mg2}

A sample of \ion{Mg}{ii} profiles starting from day 1.3 after
discovery is given in Fig. \ref{fig:prof1}a. A complete
gallery of the \ion{Mg}{ii} profiles is shown in Shore et al. (1993).

In the first IUE high resolution spectrum of day 1.3, this line
has a marked P Cygni profile consisting of  strong emission and of
very broad, shallow, and short-ward shifted absorption extending to
an edge velocity  \vedge = ~3360 km~s$^{-1}$.  The broad absorption
component indicates that there is a large velocity gradient in the
column to the observer.

The equivalent width of the emission component on day 1.3 (about 65
\AA) exceeds that of the absorption by about 30\%. This indicates that
the line is not formed by pure resonance scattering, but shows the
presence of an extra source of excitation by electron collisions
(Castor \& Lamers 1979; Lamers, Cerruti--Sola \& Perinotto 1987). This
mechanism has been already suggested for the interpretation of the
\ion{Mg}{ii} profile in V1668 Cyg 1978 (Cassatella et al. 1979).

The evolution of the \ion{Mg}{ii} line during the first two weeks can
be described as a change from a P Cygni profile at day 1.3 into a pure
emission profile at day 6 and later.
The intensity ratio of the emission with respect to
the absorption component increases from about 1.3 in the spectrum of
day 1 to about 6 on day 14.  The strengthening of the \ion{Mg}{ii}
emission continues until day 21, when a maximum value is reached
(see Fig. \ref{fig:lowres}).

The broad absorption component of the P Cygni profile decreases in
strength with time until, by day 16, it has vanished. When it gets
weaker, it unveils the presence of much narrower violet--shifted
absorption lines, which can be identified as belonging to the
principal absorption system.  (These lines are possibly present also
before day 16, but cannot be detected due to the depth of the P Cygni
absorption component.) The radial velocity of these components
increases with time (see Fig.  \ref{fig:prof1}a) reaching -2020
km~s$^{-1}$ on day 135, the last time these lines are seen (see
Sect. \ref{sec:sistemas1}). Note that the diffuse--enhanced components
are not seen in \ion{Mg}{ii} (see Sect. \ref{sec:sistemas}).

From day 53 to day 211 one can detect, superimposed to the
\ion{Mg}{ii} emission, narrow absorption components from the same
doublet at a {\it constant} radial velocity of about -600 \kms (see
Fig. \ref{fig:prof1}a).  These components will be discussed  in
Sect. \ref{sec:narrow}.

On day 192, the \ion{Mg}{ii} line appears blended with the
\ion{[Mg}{v]} 2783 \AA\ emission, which on day 296 reaches about half
the intensity of \ion{Mg}{ii} (see Fig. \ref{fig:prof1}a).  By this
time the nova is far into the  nebular phase.  (The \ion{[Mg}{v]}
emission may be present at earlier times, but is too faint to be
detected.)

%**********************************************************************
\subsubsection{\ion{S}{ii} uv1 and \ion{Si}{ii}  uv4}

The \ion{S}{ii} uv1 and \ion{Si}{ii} uv4 transitions appear as
emission--absorption features in the high resolution early spectra
after day 16, when they start becoming detectable. As shown in
Fig. \ref{fig:prof1}b, only the principal system is present in these
low excitation lines.

\subsubsection{\ion{C}{ii} 1334.53, 1335.68 \AA} 
The \ion{C}{ii} resonance doublet in Fig. \ref{fig:prof1}c shows a
marked P Cygni profile, although it is barely recognized in the
earliest high resolution spectra due to the poor signal level and
the heavy line blending.  The emission component, like that of
\ion{Mg}{ii}, is stronger than the absorption, and is flanked by
narrow overlying absorptions which nearly disappear after day
35--40. The components from the \ion{C}{ii} principal and
diffuse--enhanced systems are both detectable until day 53, but their
radial velocities cannot be determined accurately because of the small
separation of the doublet lines (the radial velocities indicated in
Fig. \ref{fig:prof1}c are merely indicative and will not be used
elsewhere).

\subsubsection{\ion{O}{i} 1302.17, 1304.86, 1306.02 \AA} 
The evolution of the \ion{O}{i} uv1 resonance triplet is shown in
Fig. \ref{fig:prof1}d. This line is blended with the \ion{Si}{ii} uv3
multiplet.  Note the broad and asymmetric shape in the earliest
spectra, and the apparent absence of the diffuse--enhanced absorptions
from both ions. { Since this line is formed by fluorescence (see
Sect. \ref{sec:observa}), its profile cannot be considered as a
classical P Cygni profile.

\subsubsection{\ion{Al}{iii}  1854.72, 1862.78 \AA}
As shown in Fig. \ref{fig:prof2}a, the behaviour of the \ion{Al}{iii} uv1 
doublet is in some respects similar to that of \ion{Mg}{ii}: its
shortward shifted absorption on day 1.3 has about the same edge
velocity ($\vedge\ \approx $ 3150 km~s$^{-1}$) as \ion{Mg}{ii}; the
strength of the emission exceeds that of the absorption, their ratio
being an increasing function of time; it appears as a strong feature
already in the earliest spectra (this property is common to other Neon
novae, see Shore et al. 1994).

The first radial velocity measurements of the principal system become
possible on day 6.3, and give a value of about - 1200 km~s$^{-1}$
(note that the principal system components were not detectable before
day 12.3 in \ion{Mg}{ii}).  Later on, on day 34, due to the very
strong and wide absorption, the higher velocity components from the
diffuse--enhanced system also appear, at a radial velocity of about
-2600 km~s$^{-1}$.  The \ion{Al}{iii} diffuse--enhanced absorption
system is clearly present until day 60, but is certainly absent in the
following observation of day 80.

\subsubsection{\ion{Si}{iv} 1393.76, 1402.77 \AA}

As shown in Fig. \ref{fig:prof2}b, the 1400 \AA\ region is 
crowded with absorption lines. The shortward shifted absorption
components from the principal and the diffuse--enhanced absorption
systems of the \ion{Si}{iv} resonance doublet begin to be seen clearly
on day 6.3.  Unlike \ion{Mg}{ii} and \ion{Al}{iii}, the P Cygni
emission component of \ion{Si}{iv} does not show up in the earliest
spectra, due to the strong overlying absorption, but starts to emerge
around day 23, and becomes quite strong on day 53.  The
diffuse--enhanced system is absent after day 60, the same as for the
\ion{Al}{iii} line.

\subsubsection{\ion{C}{iv} 1548.20, 1550.77 \AA}

The \ion{C}{iv} resonance doublet is a very interesting feature for
the way it gradually emerges and eventually shows up clearly in the
spectra due to the decrease of overlying absorption by \ion{Fe}{ii}
lines.  Its evolution is shown in Fig. \ref{fig:prof2}c.  For a
detailed identification of the \ion{Fe}{ii} lines on day 28 see
Fig. \ref{fig:prof4}.  One can easily appreciate that the 1550 \AA\
region is strongly blanketed by overlying absorption lines from the
principal and diffuse--enhanced systems of \ion{Fe}{ii} uv44 and uv45,
and of \ion{Si}{ii} uv2 (see also Sect. \ref{sec:sistemas1}).  The
line blending and the degree of saturation is so severe that the
identifications are very difficult in the earliest spectra (see for
example the spectrum of day 12 shown in the top panel of the
figure). Despite  this, it is possible to recognize the shortward
shifted absorption trough of the \ion{C}{iv} line already on day 12
and 28 (marked with an arrow in the latter spectrum). The complete
sequence of ${\it IUE}$ spectra, which cannot be shown here, indicates
a progressive ``clearing'' of the overlying absorption until around
day 40, when the P Cygni characteristics of \ion{C}{iv} become
evident.  The overlying absorptions from \ion{Fe}{ii} multiplets uv44
and uv45 become drastically reduced when, sometime after day 53 and
before day 60, the components from the diffuse--enhanced absorption
system totally disappear. By day 146 the \ion{C}{iv} feature has
narrowed considerably and the \ion{[Ne}{iv]} 1600 \AA\ forbidden line
has become comparatively stronger, as shown in
Fig. \ref{fig:prof2}c. The \ion{[Ne}{iv]} line reaches the maximum flux
on day 173.

%**********************************************************************

%**********************************************************************

\subsubsection{\ion{N}{v} 1238.80, 1242.78 \AA}
\label{sec:n1240}

The profile of the \ion{N}{v} \AA\ resonance doublet is shown in
Fig. \ref{fig:prof2}d together with that of hydrogen Lyman
${\alpha}$. The latter line shows a broad central absorption dip
of interstellar origin, discussed by Shore et al. (1993).  The
\ion{N}{v} line begins to be clearly seen around day 47. By day 123
the short-ward shifted components of the principal system become
measurable.  Their position are indicated in the figure (note that the
diffuse--enhanced components are not expected for this late date).

On day 123 one also starts  to detect the presence of  narrow
shortward shifted absorption components of the \ion{N}{v} doublet at a
radial velocity of ${\approx}$ -600 km~s$^{-1}$, superimposed on the
broad emission. Contrary to the principal absorption lines, these
components, which are indicated in Fig. \ref{fig:prof2}d for day 192,
remain {\it stable} in wavelength until the last good quality short
wavelength high resolution observations of day 201. See Sect. \ref{sec:narrow}
for a further discussion.
 
%**********************************************************************

\subsection{The widths of the emission lines}
\label{sec:widths}
\begin{figure}
\begin{center}
\psfig{file=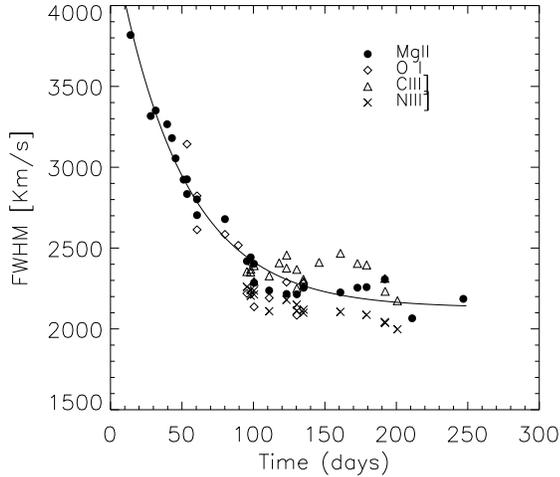,width=8cm}
\caption{The width (FWHM) of some strong emission lines  as
a function of time. The full line represents an exponential fit to the
\ion{Mg}{ii} data. The line widths were obtained by fitting the line
profiles with a Gaussian function (the absorption components
in  the permitted lines were not included)}
\label{fig:fwhm} 
\end{center}
\end{figure}
%-----------------------------------------------------------------------

We have measured the full width at half maximum (FWHM) of some of
the strongest emission lines: \ion{Mg}{ii} 2800 \AA, \ion{O}{i} 1300
\AA, \ion{C}{iii]} 1909 \AA, and \ion{N}{iii} 1750 \AA. The measured
values, ${\it not~ corrected}$ for the multiplet separation, are
plotted as a function of time in Fig. \ref{fig:fwhm} together with an
exponential fit to the \ion{Mg}{ii} data.  The figure shows an initial
rapid decrease of the line widths from ${\approx}$ 3800 km~s$^{-1}$ in
the first days, to ${\approx}$ 2450 km~s$^{-1}$ on day 90. After this
last date, the curve enters  a near--plateau region characterized
by a mean value FWHM $\approx$ 2140 km~s$^{-1}$.  It is not possible to
interpret this value in terms of expansion velocity of the nebular
region because the multiplet components cannot be separated due to the
large velocity field.   Still, the exponential decrease of the emission
line widths is a significant observational result in itself because it
suggests formation in the inner and denser regions of the wind,
where the expansion velocity is smaller.

We note that the width of the \ion{C}{iii]} 1909 \AA\ doublet is not
easy to evaluate because it is partially blended with the
\ion{Si}{iii]} 1893 \AA\ doublet, as may be appreciated in the
spectrum of day 80 (see Fig. \ref{fig:prof3}). Shore et al. (1993)
suggested a tentative identification of the 1893 \AA\ feature with
\ion{Fe}{iii} with possibly a contribution from \ion{C}{iii} 1895
\AA. Our main arguments in favour of the identification with
\ion{Si}{iii]} are the following: a) the line was detected for the
first time, at its nominal wavelength, on day 80, i.e. at the same
time as the \ion{C}{iii]} line (as it should, given the similar
excitation conditions); b) the flux ratio \ion{C}{iii]}/\ion{Si}{iii]}
increased with time (until the last high resolution observations on
day 173), which is the expected consequence of the decreasing electron
density in the ejecta (Nussbaumer \& Stencel 1987).
 
%**********************************************************************

\begin{figure}
\begin{center}
\psfig{file=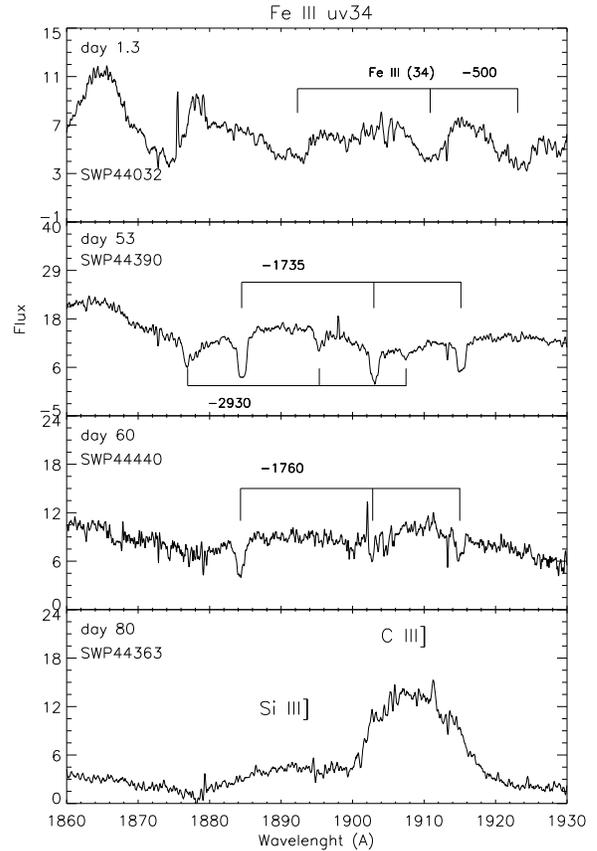,width=8cm}
\caption{Profile of the \ion{Fe}{iii} multiplet uv34 at different
dates. Note the disappearance of the diffuse--enhanced system by day
60 and the appearance of the \ion{Si}{iii]} and \ion{C}{iii]} doublets
by day 80. Fluxes are as in Fig. \ref{fig:prof1} }
\label{fig:prof3}
\end{center} 
\end{figure}
%-----------------------------------------------------------------------

\section{The absorption line systems}
\label{sec:sistemas}

\subsection{Description and evolution of the absorption line systems}
\label{sec:sistemas1}

Examples showing the identification of the principal system and
of the diffuse--enhanced system at different dates and in different
spectral regions are shown  in Figs. \ref{fig:prof1},
\ref{fig:prof2}, \ref{fig:prof3} and \ref{fig:prof4}.

The first {\it IUE} short wavelength high resolution spectrum was
collected on day 1.3, i.e.  2.2 days before visual maximum. Its
complexity does not allow accurate radial velocity determinations to
be made except in the 1890--1930 \AA\ region where, due to the lower
opacity, one can identify the principal system component of the
\ion{Fe}{iii} uv34 multiplet lines at 1895.45, 1914.05 and 1926.3 \AA\
at a radial velocity of roughly -500 km~s$^{-1}$ (see Fig.
\ref{fig:prof3}).

The principal system (low velocity) is present in all the permitted
transitions studied. The diffuse--enhanced system (high velocity) is
detected in the same transitions as the principal system (but only
until day 60, as mentioned before) with the exception of the zero volt
lines of the low ionization species of \ion{Mg}{ii} (see
Fig. \ref{fig:prof1}a), \ion{S}{ii} uv1, \ion{Si}{ii} uv4 (see
Fig. \ref{fig:prof1}b), and \ion{O}{i} uv1 (Fig.
\ref{fig:prof1}d). This might be due to the low signal level in the
wings of the corresponding emission line components. In any case, the
presence of the diffuse--enhanced absorption system components from
the ${\it excited}$ levels of the \ion{O}{i} 7773 \AA\ triplet and the
\ion{Mg}{ii} 7876 \AA\ doublet has been reported in the ground spectra
of V1974 Cyg obtained by Andrillat \& Houziaux (1993) during the early
outburst.  The higher excitation of the diffuse--enhanced system
compared to the principal system suggests that the diffuse--enhanced
system is formed deeper inside the nova envelope (where the excitation
conditions are higher) than the principal system.

The absorption components from the principal system  are
last seen in \ion{O}{i} and \ion{Fe}{iii} on day 60, \ion{Mg}{ii} on
day 135, and \ion {Si}{iv} on day 146.  Their disappearance is the
result of a {\it slow} process caused by the increasing level of
ionization and by the decrease of the column density due to the
expansion.

The diffuse--enhanced system lines are absent in the spectra of day
  1.3, but present in the spectra of day 6.3 and later. They have
  vanished {\it suddenly} after day 60 for all the lines in which they
  had been detected earlier. The exact time cannot be determined due
  to the observational gap between day 53 and 60.  This point is
  discussed further in Sect. \ref{sec:stratif}.

On day 42 $\pm$ 7, i.e. shortly before the disappearance of the
diffuse--enhanced system, the \ion{O}{i} emission reached a flux
maximum (see Fig. \ref{fig:lowres}). This event is a clear signature
that the envelope started to be transparent to the L $\beta$
radiation which pumps this transition (Paper I; Woodward et al. 1995),
as demonstrated by the subsequent appearance of emission lines with
increasing degree of ionization (see Sect. \ref{sec:observa}).
Shortly after that time, a broad P Cygni profile from the \ion{C}{iv}
1550 \AA\ doublet popped up in the spectrum as a consequence of the
decreased opacity in the envelope.  On day 63 hard X--ray emission was
detected for the first time with ${\it ROSAT}$ (Krautter et al. 1996;
Balman, Krautter \& \"Ogelman 1998).  Around day 70 important changes
in the polarization properties were reported (Bjorkman et al. 1994).
So we see that around day 53 - 70 significant changes occurred in the
spectrum.  These circumstances, all together, are discussed in Sect.
\ref{sec:discussion}.

%-----------------------------------------------------------------------
\begin{figure*}
\begin{center}
\psfig{file=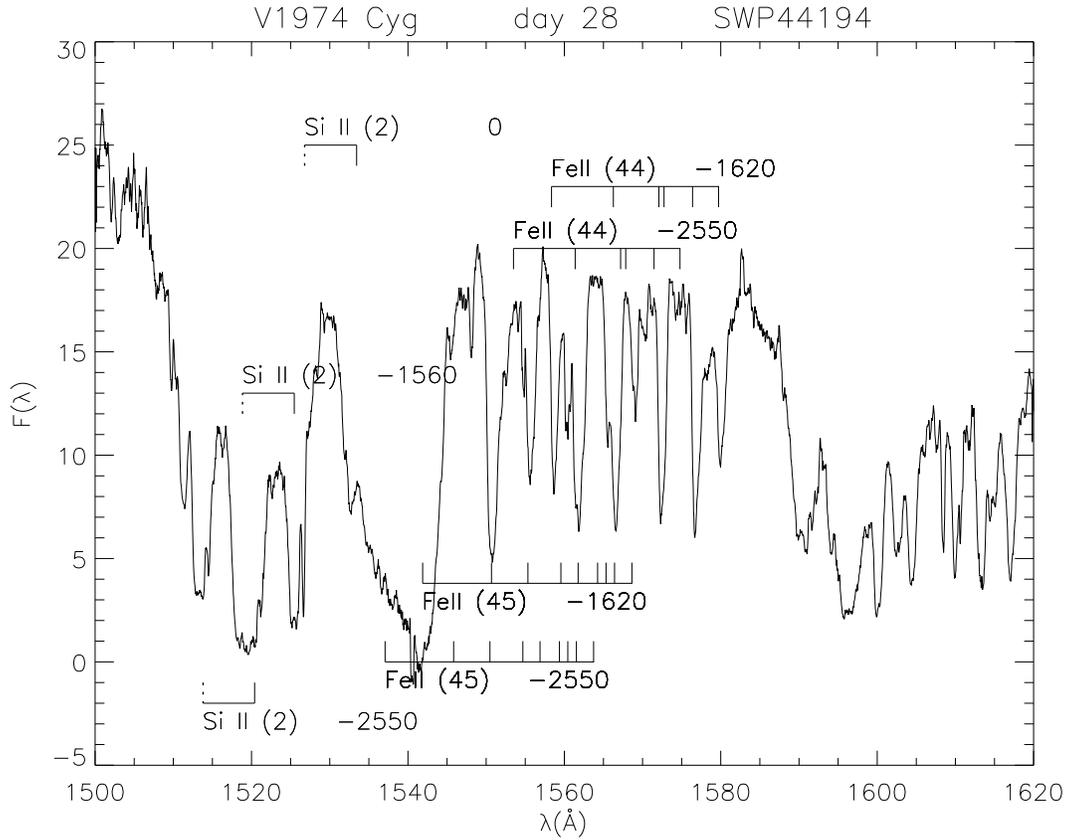,width=15cm}
\caption{
Full line identifications in the \ion{C}{iv} region on
day 28.  The figure shows how the \ion{C}{iv} emission--absorption
feature is affected by the overlying absorption from the principal and
diffuse--enhanced systems of \ion{Fe}{ii} uv44--uv45 and \ion{Si}{ii}
uv2.
The unshifted \ion{Si}{ii} lines are interstellar.  Fluxes are
in the same units as in Fig. \ref{fig:prof1} }
\label{fig:prof4} 
\end{center}
\end{figure*}
%-----------------------------------------------------------------------

We note finally that our principal system corresponds to what Chochol
et al. (1997) referred to as the spherical component from the outer
envelope. From day 2 to 62 these authors reported the presence of two
additional absorption systems in the \ion{Mg}{ii} 2800 \AA\
doublet. One of them would correspond to the range of velocities
displayed in other lines by the diffuse--enhanced system.  The other
system has a radial velocity in the range ${\approx}$ -3790 to -4580
km~s$^{-1}$.  Such high velocity absorption components have not been
found by us in any of the spectral regions analysed. After a careful
analysis of the \ion{Mg}{ii} profiles we cannot confirm the presence
of these high velocity components. We believe that these lines are due
to the diffuse--enhanced system from the \ion{Fe}{ii} uv32 lines
${\lambda}{\lambda}$ 2790.75 and 2797.04 \AA.
%**********************************************************************

\subsection{The stable absorption components}
\label{sec:narrow}

As anticipated in Sects. \ref{sec:mg2} and \ref{sec:n1240}, there is
evidence for stable and quite narrow absorption components from the
\ion{Mg}{ii} and the \ion{N}{v} doublets, which appear superimposed on
the short wavelength wings of the corresponding emission lines, as
shown in Figs. \ref{fig:prof1}a and \ref{fig:prof2}d for \ion{Mg}{ii}
and \ion{N}{v} on day 80 and 192, respectively.  The peculiarity of
these absorption lines is that their radial velocity, of about - 600
\kms, is {\it stable} to within about 20 \kms\ over a long period of
time, i.e. from day 53 (when are first seen) to day 211 (last usable
spectrum) for \ion{Mg}{ii}, and from day 123 (when are first seen) to
day 201 (last usable spectrum) for \ion{N}{v}. The Doppler widths of
these absorption lines can be measured with some accuracy only when
the underlying emission is strong enough. The values obtained are
somewhat larger for \ion{N}{v} (FWHM ${\approx}$ 180 \kms\ in days
179-201) than for \ion{Mg}{ii} (FWHM ${\approx}$ 115 \kms\ in days 95
-- 211), while the equivalent width (only measurable for the k line
components) is about the same (${\approx}$ 0.15 -- 0.20 \AA).

%**********************************************************************

\subsection{The expansion velocity of the absorption systems}
\label{sec:expansion}

%-----------------------------------------------------------------------
\begin{table*}
\begin{center}
\caption{Coefficients of the radial velocity law of the principal and
diffuse--enhanced systems in Eq. \ref{equ:veleq}}
\begin{flushleft}
\begin{tabular}{l  c  c c c c r}
\hline 
\hline 
\noalign{\smallskip}
Absorption system& $v_{\infty}$ (km~s$^{-1}$) &  $(v_{\infty} - v_0)$ (km~s$^{-1})$ & $\tau$ (days) & r.m.s. (km~s$^{-1}$)& n & Label\\ 
\hline
\noalign{\medskip}
Principal t$<$60 (days)    &  1752.31$\pm$4.42 & 1135.36$\pm$11.39& 13.139 $\pm$0.259 & ~76 &  231 & A \\  
Principal (all data)       &  1974.83$\pm$5.11 & 1102.48$\pm$~6.79& 27.580  $\pm$0.447 & ~95 &  309 & B \\  
Diffuse   (all data)       &  2898.89$\pm$8.46 & 1769.65$\pm$13.15& 16.396 $\pm$0.302 & 102 &  124 &   \\ 
 
%Principal T$<$60 (days)    &  1752.31$\pm$4.42 & 1135.36$\pm$11.39& 13.139 $\pm$0.259 & 6.5 &  242 & A \\  
%Principal (all data)       &  1974.83$\pm$5.11 & 1102.48$\pm$~6.79& 27.580  $\pm$0.447 & 10.1 &  289 & B \\  
%Diffuse   (all data)       &  2898.89$\pm$8.46 & 1769.65$\pm$13.15& 16.396 $\pm$0.302 & 11.7 &  116 &   \\  

%Diffuse   (all data)       &  2924.88$\pm$8.30 & 1769.88$\pm$12.74& 17.220 $\pm$0.307 & 12.0 &  116 &   \\  
%Principal (all data)       &  1914.89$\pm$4.88 & 1088.46$\pm$7.38& 23.253  $\pm$0.394 & 8.9 &  281 & B \\  
%
%Principal   T$<$60 (days) &  1775.81$\pm$5.41 & 1048.99$\pm$10.8& 15.45$\pm$0.365 & 5.3 &  240 & A \\
%Principal (all data)      &  1945.54$\pm$0.19 & 1045.64$\pm$0.24& 27.12$\pm$0.505 & 6.7 &  279 & B \\  
%Diffuse   (all data)       &  3002.78$\pm$0.38 & 1638.78$\pm$0.41& 21.95  $\pm$0.515 &  7.9 &  116 &   \\  
%

\noalign{\medskip}
\hline 
\end{tabular}                       
\label{tab:coeff}
\end{flushleft}             
\end{center}                        
\noindent Note: Col. 6 provides the number of measurements, inclusive
48 and 25 measurements for the principal and diffuse--enhanced
systems, respectively, obtained from optical spectra (Annuk et
al. 1993; Andrillat \& Houziaux 1993)
\end{table*}                    
%-----------------------------------------------------------------------

Let us concentrate on the radial velocity variations of the principal
and diffuse--enhanced systems in the 10 spectral regions shown in
Figs. \ref{fig:prof1}, \ref{fig:prof2}, \ref{fig:prof3} and
\ref{fig:prof4}, where the most important line identifications are
reported.

The radial velocity measurements have been carried out by fitting the
absorption profile with a Gaussian function. 
The lines studied are: \ion{O}{i} 1300 \AA, \ion{Mg}{ii} 2800 \AA,
\ion{S}{ii} uv1, \ion{Fe}{ii} uv44, \ion{Fe}{ii} uv8, \ion{Fe}{iii}
uv34, \ion{Al}{iii} uv1, \ion{Si}{iv} uv1, and \ion{N}{v}.

The radial velocity measurements are given in Tables 1 to 5 of the
Appendix.
\footnote{Tables 1 to 5 of the Appendix are only available in
electronic form at CDS via anonymous ftp to cdsarc.u-strasbg.fr
(130.79.128.5) or via http://cdsweb.u-strasbg.fr/Abstract.html}.
The measurements do
not include the \ion{C}{iv} 1550 \AA\ doublet because of the severe
line blending at early dates, and the insufficient signal level at
later dates.

%**********************************************************************

\subsubsection{Accuracy of the radial velocity measurements}
\label{sec:accuracy}

Gonz\'alez--Riestra et al. (2000) estimated the wavelength accuracy of
${\it IUE}$ high resolution spectra by measuring the positions of
interstellar lines in many well exposed {\it IUE} spectra of standard
stars, and found this to be $\pm$ 5 km~s$^{-1}$.  Since our spectra
are in general not equally well exposed as those of the standard
stars, such a test has been repeated. To this purpose we have used
the interstellar resonance lines of \ion{S}{ii} 1259.53 \AA,
\ion{C}{ii} 1335.68 \AA, and \ion{Mg}{ii} 2796.35 \AA, and obtained
the following values: -13 $\pm$ 10 (20 measurements), -9 $\pm$ 6
km~s$^{-1}$ (31 measurements) and -20 $\pm$ 5 km~s$^{-1}$ (46
measurements), respectively.  The weighted mean of these measurements,
-15 km~s$^{-1}$, is taken to represent the accuracy of the relative
radial velocity of the nova with respect to the interstellar medium,
and the error $\pm$ 5 km~s$^{-1}$ (the same as found by
Gonz\'alez--Riestra et al. 2000), is the expected error in the radial
velocity measurements of individual narrow lines in our
spectra. However, we should consider that, among the absorption lines
analyzed, only the ones from \ion{Fe}{ii} uv44--uv45 and of
\ion{Fe}{iii} uv34 are about as narrow as the interstellar lines are.
Indeed, the expected error in the velocities may be up to a factor of
four larger, depending on the line width and on the signal-to-noise
ratio.

%-----------------------------------------------------------------------
\begin{figure*}
\begin{center}
\psfig{file=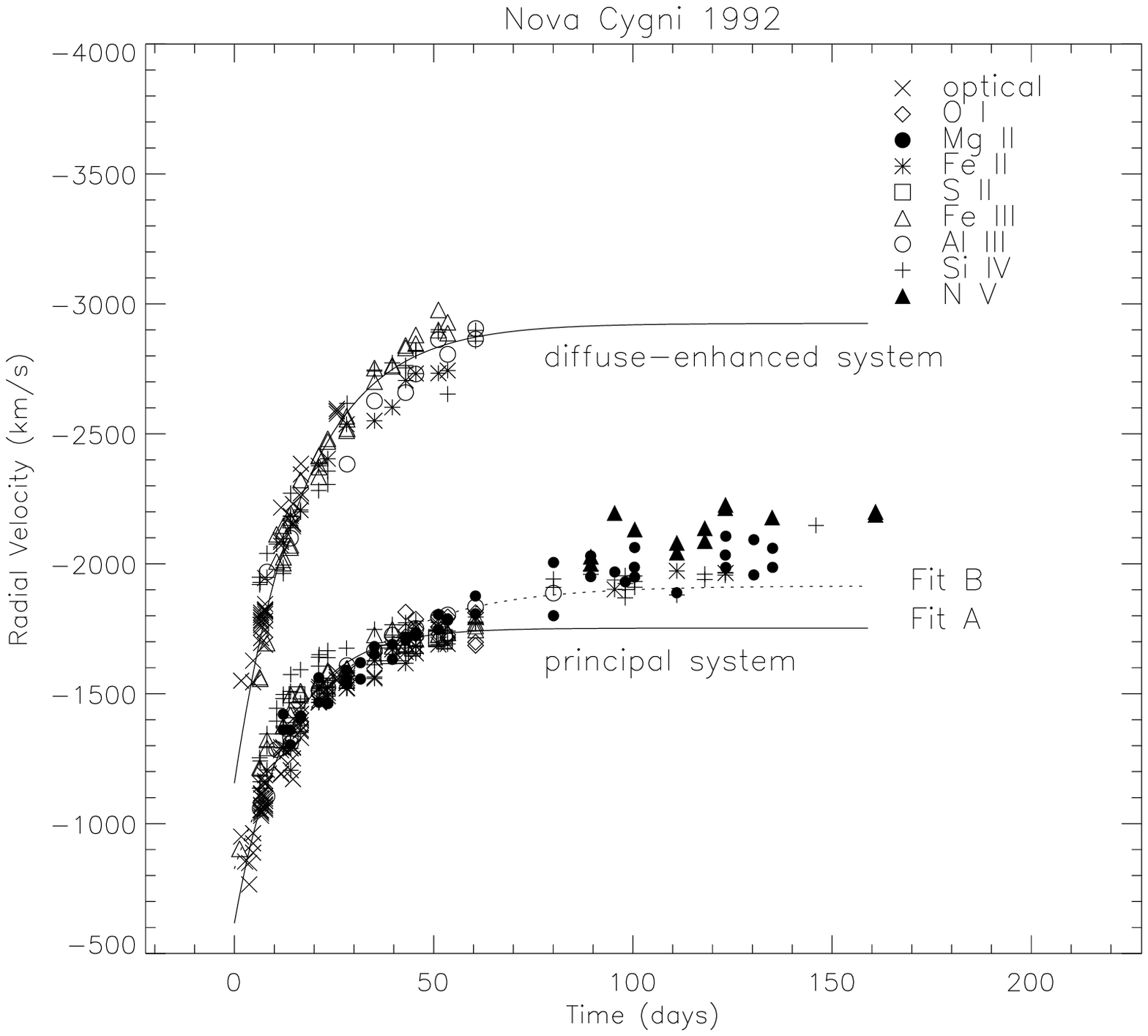,width=14cm}
\caption{Radial velocity of the principal and of the diffuse--enhanced
absorption systems for the ions indicated in the upper right part of
the figure. The velocity curves of the two systems have been fitted
with the exponential law in Eq. (\ref{equ:veleq}) to obtain the
coefficients in Table \ref{tab:coeff}.  Two fits are provided for the
principal system: $Fit~A$ is a better representation than $B$ for the
fast rising part of the curve.  Neither of the two representations is
fully satisfactory for all the observations of the principal
system. The figure suggests a sudden increase of the radial velocities
of the principal system after day day 53, with a mean value of -2000
${\pm}$ 90 km~s$^{-1}$ (49 measurements) after that date. Note the
large scatter in radial velocities after day 53}
% newvel.day.ps}
\label{fig:velocity} 
\end{center}
\end{figure*}
%-----------------------------------------------------------------------

%-----------------------------------------------------------------------
% plot dimension 15x13 cm

\begin{figure*}
\begin{center}
\psfig{file=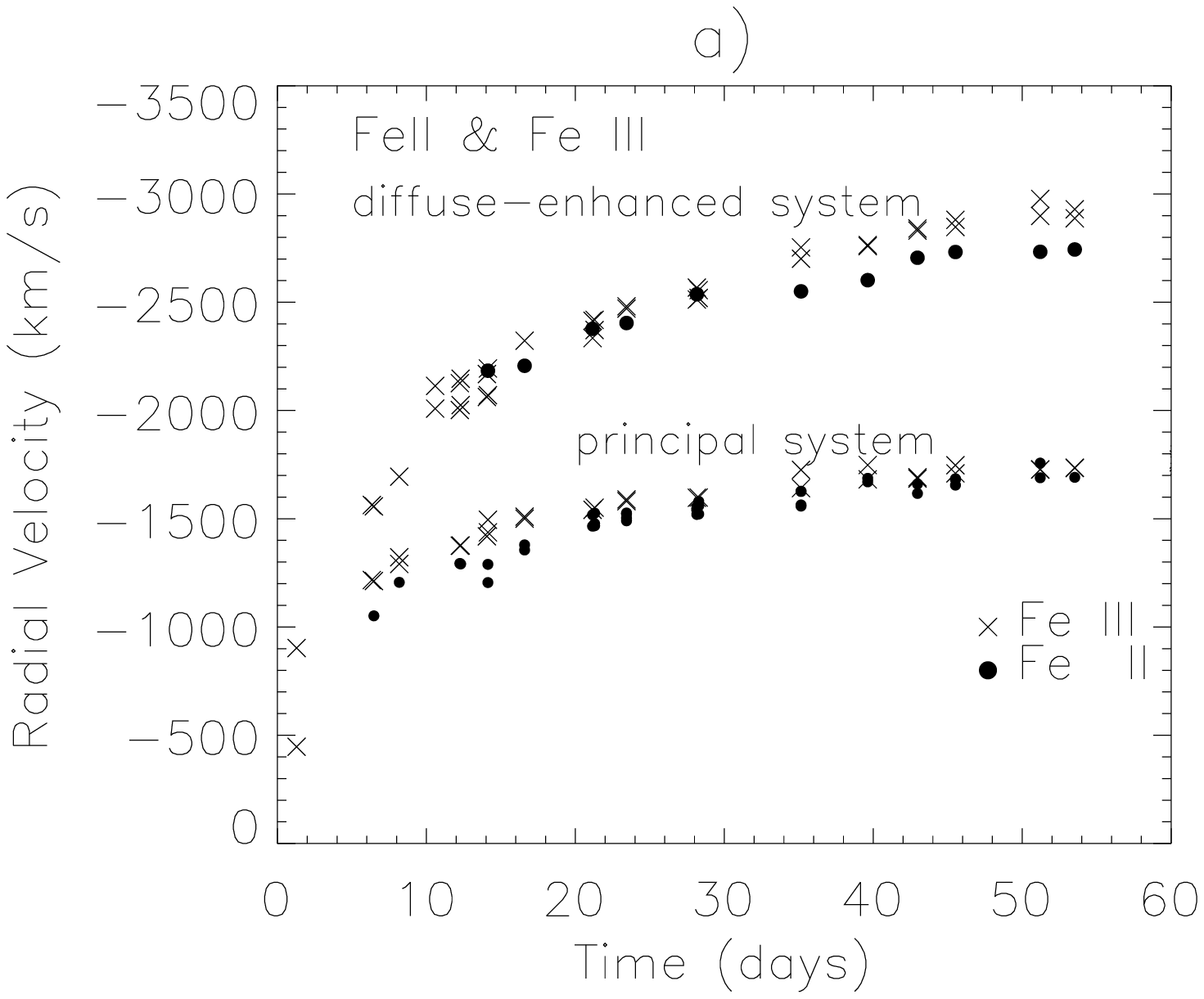,width=8cm}
\psfig{file=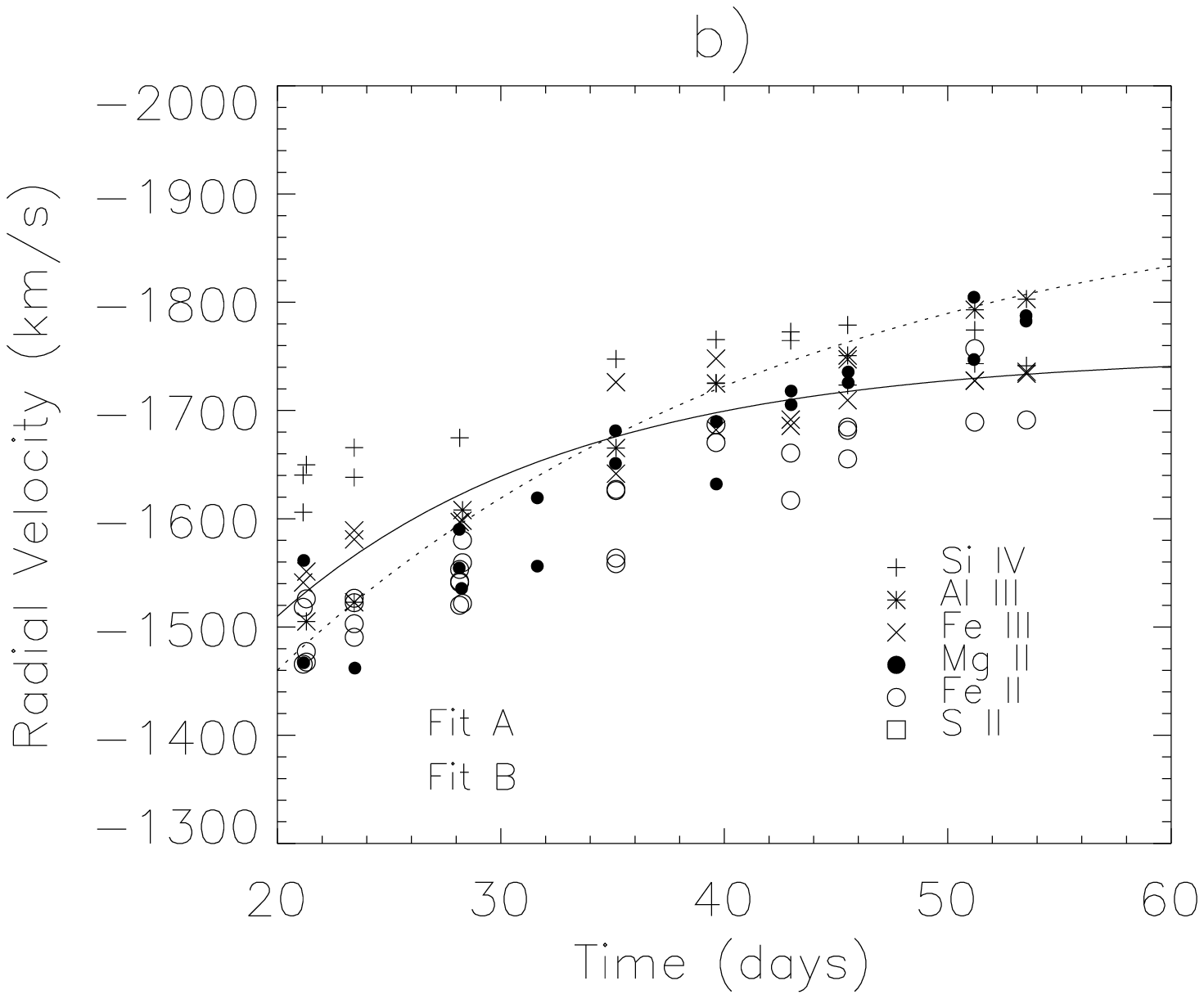,width=8cm}
\caption{Ionization stratification within the expanding regions.
Panel a) shows the radial velocities of the \ion{Fe}{ii} (dots) and
\ion{Fe}{iii} (crosses) lines from the principal and the
diffuse-enhanced systems. Panel b)  shows the  data of Fig.
\ref{fig:velocity}, but for the principal system only, and only for 
a limited range of time and velocities.  The figures indicate that
the radial velocities are systematically more negative for the higher
ionization species.}
\label{fig:stratifica} 
\end{center}
\end{figure*}
%-----------------------------------------------------------------------

%**********************************************************************

\subsubsection{The velocity law of the principal and diffuse--enhanced systems}
\label{sec:laws}

The measured radial velocities $v_{r}(t)$ of the principal and
diffuse--enhanced system are plotted in Fig.  \ref{fig:velocity} as a
function of time $t$ after the discovery.  The figure also includes
the radial velocity determinations from optical spectra (Annuk, Kolka
\& Leedj\"arv 1993; and Andrillat \& Houziaux 1993) which cover the
earliest dates, when the strongest velocity gradient is observed.

Using a non--linear least squares  algorithm,  we have fitted
the radial velocity curves with the following expression:

\begin{equation}
\it{v_r(t) = - v_{\infty} + (v_{\infty} - v_0) e^{-t/\tau}}
\label{equ:veleq}
\end{equation}

The values of the parameters so obtained are reported in Table
\ref{tab:coeff}, together with the corresponding errors, the
r.m.s. error of the fit and the number of measurements.

If the observed radial velocities $v_r(t)$ reflect an outward
expansion at a velocity $v(t)$ = -$v_r(t)$, $\tau$ ~ can be
interpreted as a characteristic time related to the acceleration of
the expanding matter, $v_{\infty}$ represents the asymptotic
(i.e. terminal) velocity of the outflowing matter, and $v_0$ is the
velocity at $t = 0$.  

Fig. \ref{fig:velocity} shows that
the radial velocity of the diffuse--enhanced system until day 60
(when it definitively disappears) is fairly well represented by
Eq. (\ref{equ:veleq}) with the coefficients in Table \ref{tab:coeff}.

The situation is more complex for the principal system, due to the
impossibility to represent with a unique fit both the rising part of
the curve (corresponding to the earliest days) and the velocity
``plateau'' starting after day 80, when the radial velocity curve
becomes rather flat.  We have therefore made two fits. The first, $
Fit~A$, was obtained from the observations made until day 60
(J.D. 2448752); it fairly well represents this period, but it
 grossly underestimates the radial velocities in the ``plateau''
region which starts after day 80.  The second, $ Fit~ B$, which
includes all the measurements  for the principal system, is a
poorer representation than $Fit~ A$ for the earliest days, when the
largest velocity variations are seen and, in addition, it clearly
underestimates the observed values after day 80.  Both fits are shown
in Fig. \ref{fig:velocity}.

The failure to represent the radial velocity of the principal system
with a unique law may be due to the limitation of the empirical
representation of Eq. (\ref{equ:veleq}).  An interesting alternative,
which we adopt as working hypothesis, is that the velocity law of the
principal system has actually changed after day 60. A careful
inspection of the data in Fig. \ref{fig:velocity} suggests, in fact,
that the radial velocity of the principal system has suffered from some
sort of discontinuity after day 60, jumping from -1760 ${\pm}$ 47
\kms\ in days 53--60, to -2000 ${\pm}$ 90 \kms\ in the ``plateau''
phase after day 80 (from 27 and 49 measurements, respectively).  In
the following we will indicate with $v^{ps}_{\rm plat}$ = 2000 \kms\
the observed mean value of the expansion velocity of the principal
system lines during the ``plateau'' phase.

As for the expansion velocity just before the jump, hereafter
indicated with $v^{ps}_{\infty}$, we have adopted the above observed
value of 1760 \kms\ instead of the asymptotic value of 1752 \kms\
derived from fitting the radial velocity data of the principal system
before day 60 (see Table \ref{tab:coeff}, Fit~$A$). The velocity jump
suffered by the principal system is then ${\Delta}v$ = $v^{ps}_{\rm plat}$ -
$v^{ps}_{\infty}$ $\approx$  240 \kms. These values, together with the
plateau velocity of the diffuse--enhanced system $v^{ds}_{\infty}$ = 2899
km~s$^{-1}$ (see Table \ref{tab:coeff}) will be used later in
Sect. \ref{sec:interpretation}.

%**********************************************************************

%-----------------------------------------------------------------------
% velocity di tutte le righe dimensione 17x16
\begin{figure*}
\begin{center}
%*******************FIG 9
%\psfig{file=fe3fe2ew.ps,width=8cm}
\psfig{file=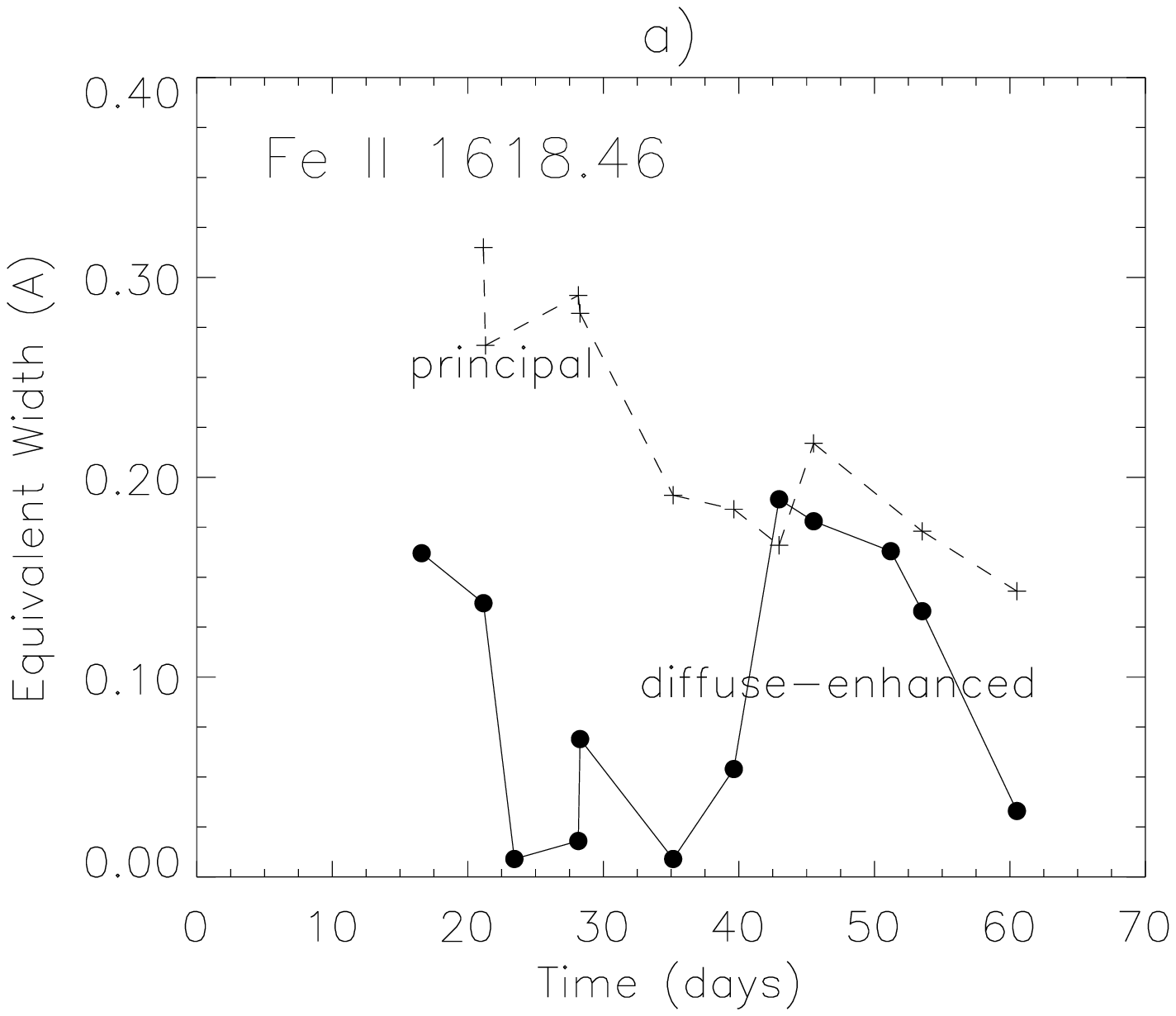,width=8cm}
\psfig{file=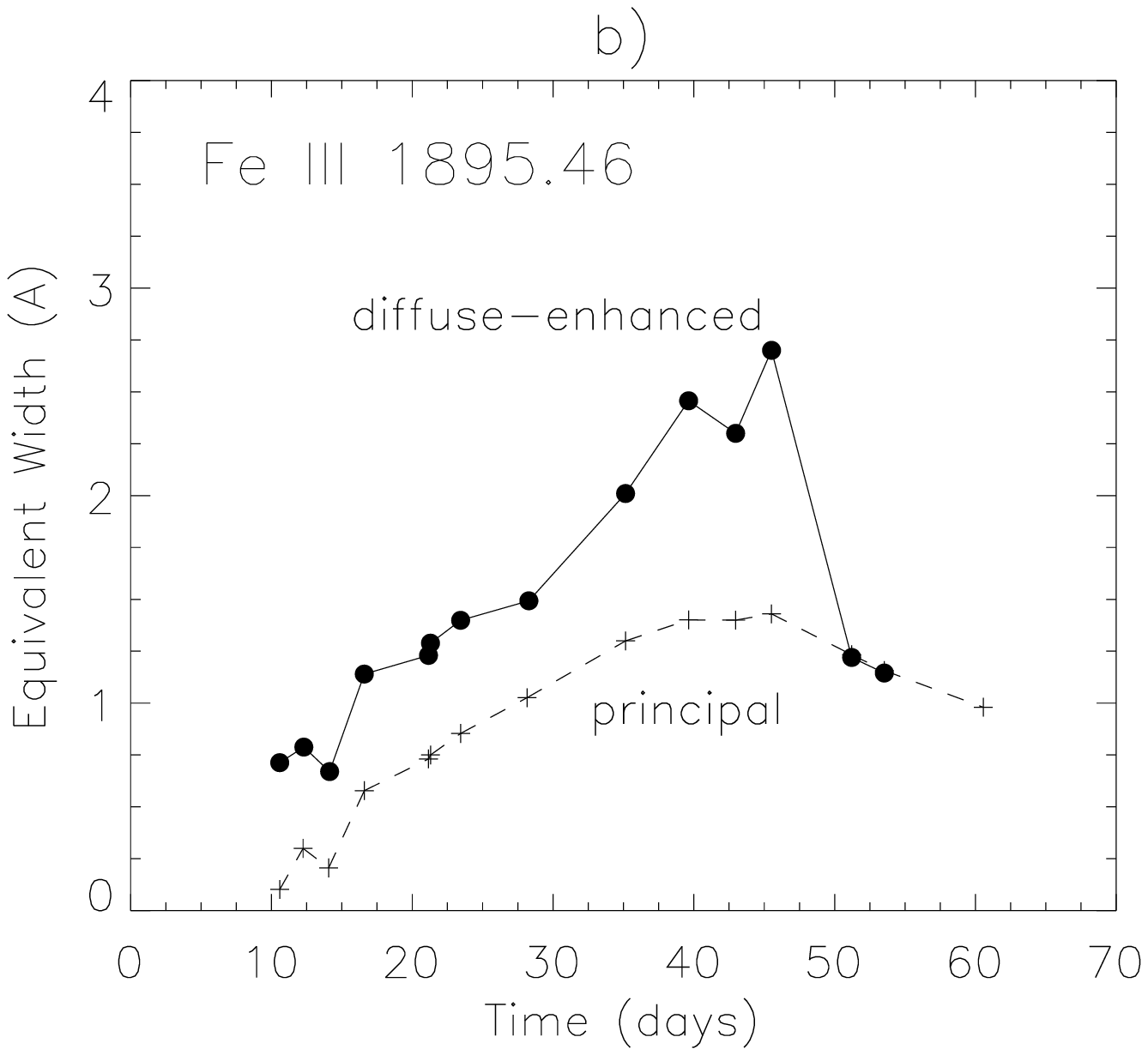,width=8cm}
\caption{The equivalent width of the absorption components from the
principal and the diffuse--enhanced system as a function of time after
outburst for: a) \ion{Fe}{ii} 1618.46 \AA\ and, b) \ion{Fe}{iii}
1895.46 \AA.  Up to day 40, the principal lines of Fe II are stronger
than the diffuse--enhanced lines, whereas the reverse is true for the
Fe III line (see Sect. \ref{sec:stratif}). Note the sudden drop of the
equivalent width for the \ion{Fe}{iii} diffuse--enhanced system on
days 51--53.  After day 53 the diffuse--enhanced component of the
\ion{Fe}{iii} 1895.46 \AA\ absorption is no longer seen
(Sect. \ref{sec:sistemas}).}
\label{fig:ferroew} 
\end{center}
\end{figure*}

\begin{figure*}
\begin{center}
\psfig{file=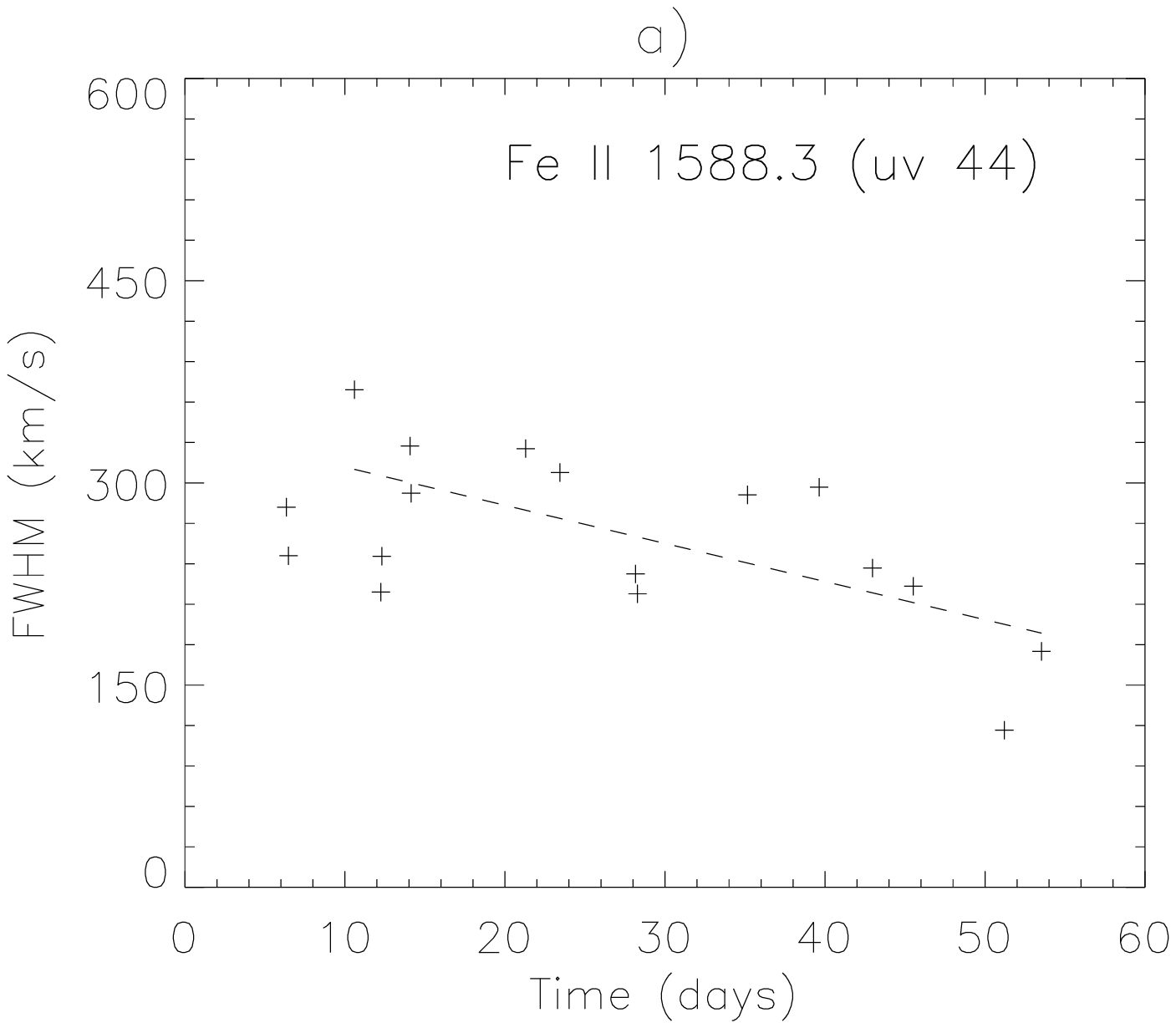,width=8cm}
\psfig{file=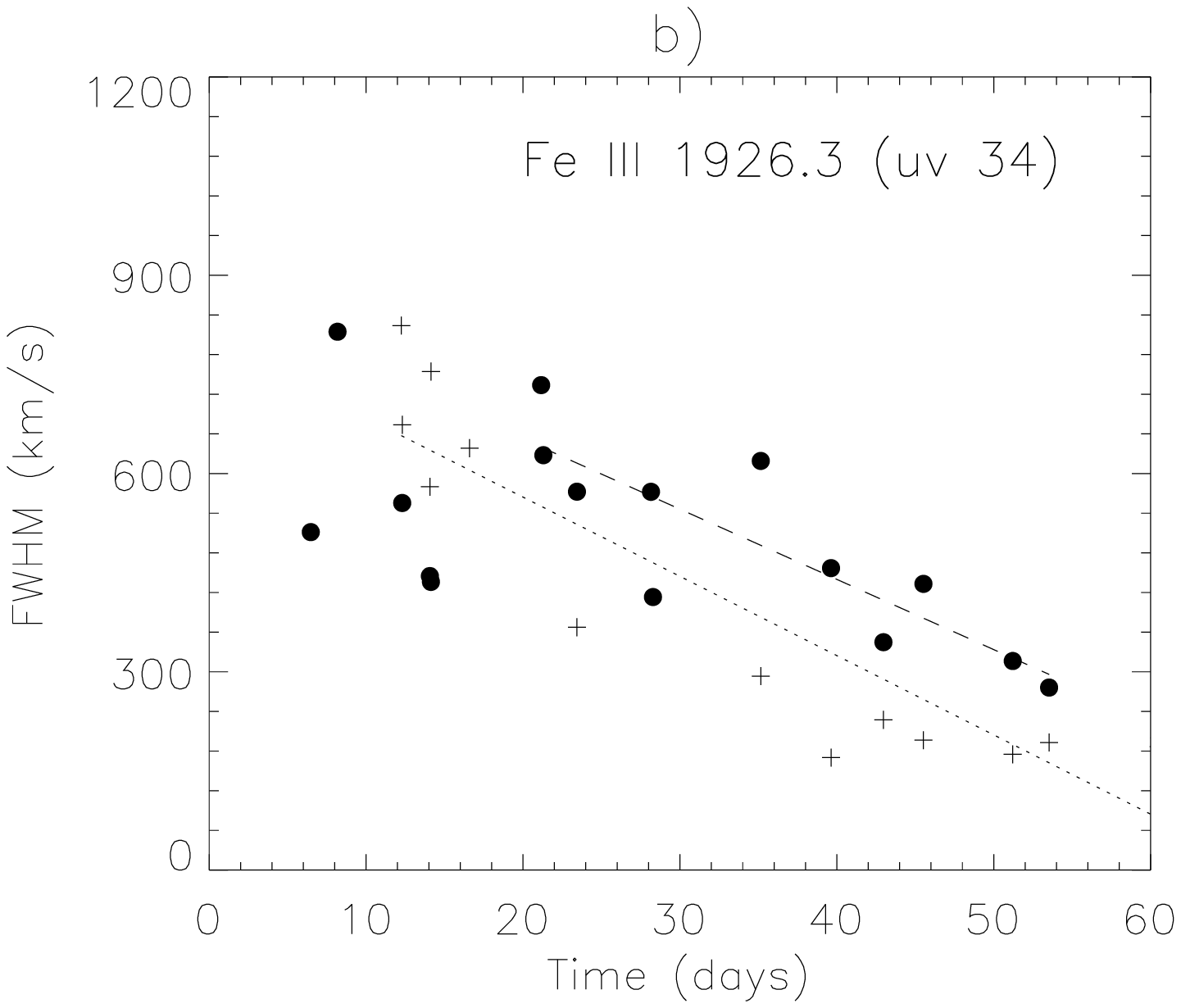,width=8cm}
\caption{The Doppler width (FWHM) of the absorption components of a)
\ion{Fe}{ii} 1588.3 \AA\ (principal system), and b) of \ion{Fe}{iii}
1926.3 \AA\ (principal and diffuse--enhanced systems) are plotted
against time after the outburst. Linear regressions to the data of the
principal system (crosses) and of the diffuse--enhanced system (dots)
are indicated with a dashed line and a dotted line, respectively}
\label{fig:fwhm_fe2}
\end{center}
\end{figure*}

%**********************************************************************

\subsubsection{Ionization stratification within the expanding envelope}
\label{sec:stratif}

As shown in Sect. \ref{sec:pcygni}, the high ionization resonance
lines, such as the \ion{C}{iv} doublet, must be formed in regions
which are deeper inside the envelope than the \ion{Fe}{ii} lines,
because the \ion{Fe}{ii} lines absorb the broad \ion{C}{iv} profile.

There is a second argument which throws light on  another aspect of the
stratification: radial velocities.  Let us consider the radial
velocity of lines of the same element, but in different ionization
stages, such as \ion{Fe}{ii} and \ion{Fe}{iii}.  The following lines,
identified Figs. \ref{fig:prof2}c and \ref{fig:prof3}, were selected
as being sufficiently strong and narrow:

\begin{itemize}
\item Principal System: 1584.95 and 1588.30 (\ion{Fe}{ii} uv44);
1608.45 and 1618.46 (\ion{Fe}{ii} uv8); 1895.46 and 1926.30
(\ion{Fe}{iii} uv34).
\item Diffuse--Enhanced system: 1895.46 and 1926.30 from \ion{Fe}{iii} 
uv34 and 1608.446 from \ion{Fe}{ii} uv8.  
\end{itemize}

The radial velocity curves of the \ion{Fe}{ii} and \ion{Fe}{iii}
lines shown in  Fig. \ref{fig:stratifica}a
not only confirm the presence of two expanding zones with different
velocity regimes, but contain information about how the ionization is
stratified within each region.  In fact, it can be clearly seen 
that the \ion{Fe}{iii} radial velocities are systematically more
negative than those of the lower excitation \ion{Fe}{ii} lines. If we
 assume that the ionization temperature decreases outward, as is well
justified both by nova models (Kato \& Hachisu 1994, Bath \& Harkness
1989) and observations (Friedjung 1966b) it follows that, at a given
time, {\it the expansion velocity decreases outward within each
region.}  Note that such stratification effects were also present in
the data of Fig. \ref{fig:velocity}. They become very clear when
plotting the radial velocity curve of the principal system in a
limited range of time and velocities, as done in
Fig. \ref{fig:stratifica}b.

The presence of a velocity stratification within each of the expanding
regions is what actually might cause the velocity spread at a given
date (${\approx}$ 70 km~s$^{-1}$ for the principal system) to be
substantially larger than the measurement errors (see
Sect. \ref{sec:accuracy}).

A third way to study the ionization structure of the envelope
is to analyse how the equivalent widths of the absorption systems vary
with time for two transitions of the same element, but in different
ionization stages.  The results of this comparison are shown in Fig.
\ref{fig:ferroew} for the lines of a) \ion{Fe}{ii} the 1618.46
\AA, and b) \ion{Fe}{iii} 1895.46 \AA.  If we restrict ourselves to
the results before day 40, we see that the \ion{Fe}{ii} line is
significantly stronger in the principal system than in the
diffuse--enhanced system while, on the contrary, the higher ionization
\ion{Fe}{iii} line is stronger in the diffuse--enhanced system.  These
observations are clearly consistent with a model in which the
diffuse--enhanced system is formed deeper inside the expanding
envelope of the nova, where the degree of ionization is higher.

Finally, the stratification effects may be seen also in the Doppler
widths (FWHM) of the \ion{Fe}{ii} and \ion{Fe}{iii} lines from the
principal and the diffuse systems.  In Fig. \ref{fig:fwhm_fe2} we show
as a function of time the width of the \ion{Fe}{ii} 1588.3 \AA\ line
from the principal system (the corresponding diffuse component cannot
be used because of blending), and that of \ion{Fe}{iii} 1926.3 \AA\
from both the principal and diffuse systems. Indicative values of the
line widths for the principal system component of \ion{Fe}{ii} 1588.3
\AA\ are: 300 and 180 \kms\ for days 15 and 55, respectively. In the
same period, the widths of the \ion{Fe}{iii} 1926.3 \AA\ line
decreases from ${\approx}$ 705 to 280 \kms\ in the diffuse system and
from ${\approx}$ 625 to 144 \kms\ in the principal system. It can be
clearly seen that a) the \ion{Fe}{iii} line is almost a factor of two
broader than \ion{Fe}{ii} and b) the width of the diffuse--enhanced
component of \ion{Fe}{iii} 1926.3 \AA\ is systematically larger than
that of the principal system.

%**********************************************************************

\section{Possible interpretations of the radial velocity curves}
\label{sec:interpretation}

\subsection{Non--spherical geometry of the outburst}
\label{sec:spherical}
The case of non--spherical geometry of the ejecta has been
investigated by Lloyd et al. (1993) to explain the increasing
observational evidence for a bipolar structure in recurrent novae.  In
this framework, the high and low velocity absorption systems could
arise from material ejected in the equatorial plane and along the
polar axis, at different speeds. 
There is indeed unconfutable evidence for departures from spherical
symmetry in the ejecta of V1974 Cyg 2--3 years after the outburst
(Paresce et al. 1995;  Shore et al. 1997).  However, should such
geometrical effects be applicable to novae in general, the wide spread
of possible system inclination angles would cause the ratio between
the expansion velocity of the diffuse--enhanced system and that of the
principal system to vary in a wide range for different novae, in
contrast with the observational evidence that it varies in a small
range of values between 1.5 and 2 (McLaughlin 1960). In V1974 Cyg this
ratio is about 1.6. 
For V1974 Cyg the major argument against a bi-directional
interpretation of the two absorption line systems comes from the fact
that both velocity systems are seen in absorption. This implies that
the absorbing matter of both systems is in the line of sight from the
emitting region (i.e. the close to the star) to the observer.  This is
difficult to reconcile with a model in which one of the systems comes
from a disk and the other from polar ejections, unless both cover a
large solid angle.

%**********************************************************************

\subsection{Discrete  shell ejections}
\label{sec:impulsive}

We propose a new model, described in detail in the next section,
in which the principal and the diffuse--enhanced systems arise from
{\it two} discrete expanding shells ejected in separate events.
The principal system shell, or main shell, is impulsively ejected at
the moment of the outburst, is located in the outermost expanding
regions, and contains the bulk of the ejected matter. A following less
massive impulsive ejection event gives rise to a second faster
expanding shell, where the diffuse--enhanced system is formed.
We provide below two arguments in favour of the above interpretation,
which come from a constraint on the angular size of the nova on day
10, and from the detection of X--ray emission on day 63.\\

\noindent a) The angular size and the distance of V1974 Cyg\\
%\label{sec:distance}

\noindent According to the interferometric measurements of Quirrenbach et
al. (1993) the angular radius of V1974 Cyg in hydrogen H${\alpha}$ on
day 10 was 2.5 ${\pm}$ 0.12 mas. The distance to V1974 Cyg can be
obtained by comparing this angular size determination with the
distance reached by the principal system expanding shell, as predicted
from the expansion law in Eq. (\ref{equ:veleq}).

The mean  radius of the shell at time $t_{10}$ = 10 days can be
obtained by integrating Eq. (\ref{equ:veleq}) over time. Using the
appropriate constants for the principal system ($Fit~ A$ from Table
\ref{tab:coeff}), and assuming that at the time of the outburst $t_0$
the radius of the shell is about 100 $R_{\sun}$, one finds $r(t_{10})$
= 1.1 10$^9$ km.  This value, combined with the angular radius in
hydrogen H${\alpha}$ on day 10 from Quirrenbach et al. (1993), leads
to a distance of 2.9 ${\pm}$ 0.2 kpc, which is in good agreement with
the distance of 2.3 to 2.9 kpc obtained by these authors by assuming a
constant velocity in the range 1000 to 1300 km~s$^{-1}$. Our
determination is also in agreement with the range of values,
${\approx}$ 2.2 -- 3.4 kpc obtained by Della Valle \& Livio (1995)
from their upgraded maximum magnitude versus rate of decline
relationship. We conclude that the velocity law of the principal
system (see Eq. (\ref{equ:veleq}) and the coefficients of $Fit~ A$ in
Table \ref{tab:coeff}) is consistent with the extended source observed
by Quirrenbach et al. (1993) on day 10, if the principal absorption
line system is formed in the outermost shell that contains most of the
nova ejecta.

Our distance determination agrees also with the distance of 1.8 to 3.2
Kpc given by Paresce et al. (1995) on the basis of FOC ${\it HST}$
imaging in 1993--1995.  According to Paresce et al. (1995) there is 
clear evidence from FOC observations that the expanding nebula was
spherically symmetric in May 1993 and that it developed an elliptical
ring shape by January 1994.  Paresce et al. (1995) interpreted this
morphological transformation, as well as the different expansion rates
along the major and minor axes in terms of deposition of angular
momentum by the 1.95 hr period binary into the initially spherically
symmetric nebula.  The presence of an expanding ring around V1974 Cyg
is also confirmed by the optical spatially resolved spectra obtained
by Rosino et al. (1996) in 1994--1995, and by GHRS ultraviolet
spatially resolved spectra obtained by Shore et al. (1997) in 1995.\\

\noindent b) The origin of the hard X--ray emission\\

\noindent Let us consider the hypothesis that the diffuse--enhanced
system arises from regions which are located deeper  inside the
nova envelope and are expanding faster than those where the principal
system is formed. We are interested in knowing if and when the fast
moving diffuse system shell will overtake the principal system shell.
This can be derived by integrating Eq. (\ref{equ:veleq}) over time.
The implicit ${\it a~ priori}$ assumption is that we are dealing with
two separate shells, i.e. that the measured velocities reflect the
velocity of the outflowing matter. 
%\label{sec:multi-phase}
 As boundary conditions we impose that, at $t_{10}$ = 10 days, the 
shell that produces the principal system has a mean radius  of about
$r(t_{10}) \simeq 1.1~10^9$ km, as derived from the interferometric
observations  (see above). The mean radius of the shell that
produces the diffuse system is not well known. 
If we assume that it is roughly 100 $\Rsun$ on day 10, and use the
relevant constants in Table \ref{tab:coeff}, we find that the diffuse
system shell will reach the principal system shell on day 30 at a
radial distance of 3.6 10$^9$ km from the white dwarf centre. In
reality, both the date of encounter and the radial distance should be
considered as lower limits, since the encounter date is mainly
determined by the low velocity in the early days, and the velocity law
in Eq. (\ref{equ:veleq}) tends to overestimate the expansion velocities
in the earliest days (see Sect. \ref{sec:laws}).  Tests in this
sense were made by making ``local'' fits (i.e. in restricted time
intervals) to the radial velocity data in Fig. \ref{fig:velocity} and
integrating the combined results. Keeping nearly the same boundary
conditions as above, we find that the encounter took place at a
later date, i.e. on day 42, at a radial distance of 5.4 10$^9$ km from
the white dwarf center.  The large changes of the equivalent width of
the diffuse--enhanced \ion{Fe}{ii} and \ion{Fe}{iii} lines which took
place after day 40 (see Fig. \ref{fig:ferroew}) suggest that the
latter date of encounter is more realistic than day 30.
We then assume that the two shells have started  to merge together
around day 42. 
Because of the finite thickness of the shells, this process will
last until the less massive diffuse system shell is fully mixed with
that of the supposedly more massive principal system. By this
mechanism it is possible to account for the disappearance of the
diffuse system between day 53 and 60, and the jump in the radial
velocity at the ``plateau'' phase of the principal system.

With this hypothesis we can actually estimate the mass ratio of the
principal and diffuse system shells $M_p$/$M_d$.  In Sect.
\ref{sec:laws} we have discussed the possibility that the principal
system shell, to reach the observed ``plateau'' value of $v^{ps}_{\rm
plat}$ = 2000 km~s$^{-1}$, has accelerated after day 53 by ${\Delta}v$
= $v^{ps}_{\rm plat}$ - $v^{ps}_{\infty}$ ${\approx}$ 240 \kms ~from
$v^{ps}_{\infty}$ = 1760 \kms.

Suppose that such an acceleration is actually the result of a merging
process of two shells with different masses and with initial and final
velocities $v^{ds}_{\infty}$ = 2899 km~s$^{-1}$ and $v^{ps}_{\rm plat}$,
respectively. We can then evaluate the mass ratio from the momentum
conservation equation as $M_p$/$M_d$ = ($v^{ds}_{\infty}$ -$v^{ps}_{\rm
plat}$)/${\Delta}v$.  With the quoted values for the velocities we
find that the principal system shell is roughly 4 times more massive
than that of the diffuse system.

If the disappearance of the diffuse--enhanced system is due to the
formation of a shock front, a hot plasma is produced, whose
temperature can be evaluated from the velocity difference
${\Delta}v_{pd}$= $v^{ds}_{\infty}$ - $v^{ps}_{\infty}$ between the
asymptotic velocities of the two systems as

\begin{equation}
T_{shock} = \frac{3}{16} \frac{ {\mu}m_H({\Delta}v_{pd})^2 }{\kappa}
\approx 12.1 ( \frac {{\Delta}v_{pd}}{100~km~s^{-1} })^2 ~~~eV
\label{equ:xtemp}
\end{equation}

\noindent
(see Lamers \& Cassinelli 1999). With $v^{ds}_{\infty}$ = 2899
km~s$^{-1}$ (see Table \ref{tab:coeff}) and $v^{ps}_{\infty}$ = 1760 \kms
(Sect. \ref{sec:laws}) 
we obtain a shock temperature $T_{shock}$ ${\approx}$ ~1.6~ keV. This value is
lower than the value of ${\approx}$ 10 and 5 keV observed by Balman et
al. (1998) on day 63 and 91, respectively, but is  close to the
nearly constant temperature of the X--ray hard emission component
reported by Balman et al. (1998) for days 97 to 653.  We can
reasonably conclude that the hard X--ray emission from V1974 Cyg is
most likely due to shock interaction of the diffuse--enhanced system
with the principal system matter.  

%***********************************************************************

\section{Towards an empirical model}
\label{sec:model}

In this section we summarize the important observational results
of this paper and  propose an empirical model to explain the
observations.

%**********************************************************************

\subsection{The observed properties}
\label{sec:observed}
\begin{enumerate}

\item The UV spectrum at early phases (day $\simless$ 14 for
\ion{Mg}{ii} and later on for higher excitation lines), is
characterized by P Cygni profiles in the major UV resonance lines,
with very high edge
velocities of about 3400 \kms. The P Cygni profiles gradually
change into full, approximately symmetric emission lines (see
Sect. \ref{sec:pcygni}).  This suggests that line formation
progressively shifts toward the inner and denser regions of the
pseudo--photosphere.

\item The width (FWHM) of the emission lines decreases rapidly with
time from about 3800 \kms\ on day 14, to 2300--2400 \kms\ on day
100. It then decreases at a much slower rate down to 2100 \kms\ on day
250.  When the emission dominates the P Cygni profiles (after about
day 20 for \ion{Mg}{ii}), the density in the wind must be high (see
Sect. \ref{sec:widths}).  So the decreasing width of the emission
components suggests that the line-emitting layers move deeper into the
accelerating layers of the wind, or that the wind velocity is
decreasing with time (but there is no evidence for this in the present
case).

\item The time of maximum emission depends strongly on the ionization
stage: Mg II reaches its maximum  on day 21, \ion{O}{i} $\lambda$
1300 \AA\ (which via fluorescence reflects the strength of Lyman
$\beta$) peaks at day 42, \ion{C}{iii]} on day 77, \ion{O}{iii} on day
143, and \ion{N}{v} on day 215 (see Sect. \ref{sec:observa}).  This
shows that the degree of ionization in the wind increases with time as
the central source gets hotter.

\item The {\it IUE} observations show that, including the very first
days, low excitation narrow absorption lines can be detected
superimposed on the P Cygni profiles or the emission profiles (see
Sect. \ref{sec:evolution}).  This indicates that the emission
lines are formed {\it inside} the region where the absorption
components are formed.

\item Two velocity systems of absorption lines can be recognized: the
{\it principal system}
and the {\it diffuse-enhanced system}.
At any time, the expansion velocity of the diffuse--enhanced
system is higher than that of the principal system by about 600 to
1000 \kms. The expansion velocity of both systems 
increases with time according to
Eq. (\ref{equ:veleq}), exponentially reaching a velocity of ${\approx}$
1750 \kms\ and ${\approx}$ 2900 \kms, respectively (see Table
\ref{tab:coeff}).

\item The principal system lines are present in the spectra of day
1.3. The diffuse--enhanced system lines are absent on day 1.3 but they
are present in the spectra of day 6.3. The optical spectra by
Andrillat \& Houziaux (1993) show that the diffuse--enhanced lines are
first seen at about day 3.3 and are already strong on day 5
(unfortunately we do not have UV-spectra in this period).  This shows
that the P Cygni profiles that are observed on day 1.3 mark a phase
{\it in between} the formation of the principal and of the
diffuse--enhanced system lines.

\item The width (FWHM) of the absorption components of both the
principal and the diffuse systems decreases with time,
but the diffuse system lines are generally broader than the principal
system lines (see Fig. \ref{fig:fwhm_fe2}).

\item Within both regions that form the principal system and the
diffuse--enhanced system lines, the higher ionization lines have a
slightly higher expansion velocity. The velocity difference is of the
order of 200 \kms\ only.  This suggests that within each of these two
regions, at any time, the higher velocity layers are deeper than the
lower velocity layers.

\item After day 60 the diffuse--enhanced system lines disappear
suddenly, and shortly thereafter the principal system lines show a 
discontinuous increase in velocity of $\approx$ 240 \kms. Around the
same period, other significant changes in the spectrum occur: hard
X-rays are observed for the first time, and the degree of polarization
changes.  This has been explained in Sect. \ref{sec:impulsive}, by
means of an encounter/collision and final merging of the two shells
where the two absorption line systems are formed.

\item From about day 53 to 211, when there is enough signal in
the short wavelength emission wing of the \ion{Mg}{ii} emission, narrow
shortward--shifted absorption components are seen superimposed on the
emission at a {\it stable} radial velocity of about -600 \kms (Fig.
\ref{fig:prof1}a).  The same stable absorption components at -600
\kms\ are also seen in the \ion{N}{v} doublet from day 123 to 201, but
are best detected near the  latest  dates, when the underlying
doublet emission has become strong enough (see Fig. \ref{fig:prof2}d).
These narrow and stable absorption lines are probably formed at a very
large distance from the star from matter ejected in the WD--wind
before the nova eruption, not yet perturbed by it.  Their later
appearance in \ion{N}{v} than in \ion{Mg}{ii} is a signature of the
increasing degree of ionization in the pre--nova wind.

\end{enumerate} 

%*****************************************************************************

\subsection{A multi-phase model}
\label{sec:multi-phase}

To explain the observed characteristics, we consider a schematic
dynamical model in which mass loss has gone trough five phases.  Each
mass loss phase is responsible for the occurrence of certain
spectroscopic features.  In a time sequence of events, these phases
are:

\begin{description}
\item{~I} The {\it pre-nova wind phase}.  A low mass loss wind from
the white dwarf sets in with a velocity of about -600 \kms\  long
before the outburst.

\item{II} The {\it main ejection phase}.  This phase probably starts
about 3 days before the discovery time (see Introduction) and lasts a
very short time ($\simless$ 1 day).  It is characterized by a high
mass loss rate, and by an initial wind velocity of at least
${\approx}$ 500 \kms at day zero. This high mass loss phase produces
the main shell that contains most of the ejected matter and gives rise
to the principal absorption line system. During this phase, the
ejection velocity increases with time, probably due to a decreasing
mass loss rate, as e.g. predicted in the optically thick nova ejection
models by Kato \& Hachisu (1994).

\item{III} An {\it intermediate wind phase} with a lower
%\footnote{HENNY: lower of previous phase?} 
mass loss rate but with a very high outflow velocity.  It is
responsible for the P Cygni profiles seen on day 1.3.

%\item{III} An {\it intermediate wind phase} with a lower mass loss
%rate but with very high velocity.  It is responsible for the P Cygni
%profiles seen on day 1.3.  

\item{IV} A {\it second ejection phase} of high mass loss rate. It
takes place at some stage between day 3 and 5
and lasts a very short time ($\simless$ 1 day).  The mass loss rate is lower
and the velocity is higher than during the first ejection phase.  Also
during this phase the wind velocity increases with time, possibly
accompanied by a decreasing mass loss rate. The resulting shell gives
rise to the absorption components of the diffuse system.

\item{V} A {\it high velocity wind phase} of relatively low mass loss
rate.  It lasts relatively long, about $10^2$ days at least. The P
Cygni profiles and the emission lines are produced in this wind.  This
phase represents the natural, smooth 
continuation of phase III, but is characterized by a lower mass loss
rate and ejection velocity.
%\footnote{HENNY: correct?}

\end{description}

The above time sequence of mass loss phases (or episodes in the case
of the main and second ejections) gives rise to a highly stratified
envelope, whose structure changes with time.  To illustrate these
changes we have sketched in Fig. \ref{fig:modello} the radial profile
in density and velocity for day 10 and day 60. The structure of the
envelope around day 10 is rather complex. From inside to outside we
find: the {\it high velocity wind} (which is accelerated in a
optically thick inner region), the {\it second shell} (giving rise to
the diffuse system), an {\it intershell region} between the second
shell and the main shell (we have little information on this shell
except that its column density must be so low that it does not produce
detectable absorption components), the massive {\it main shell}, and
the {\it pre--nova} white dwarf wind.  By day 60 the second shell has
merged with the main shell and the wind is no longer optically thick.

%-----------------------------------------------------------------------
\begin{figure*}
\vskip 0.3cm
\psfig{file=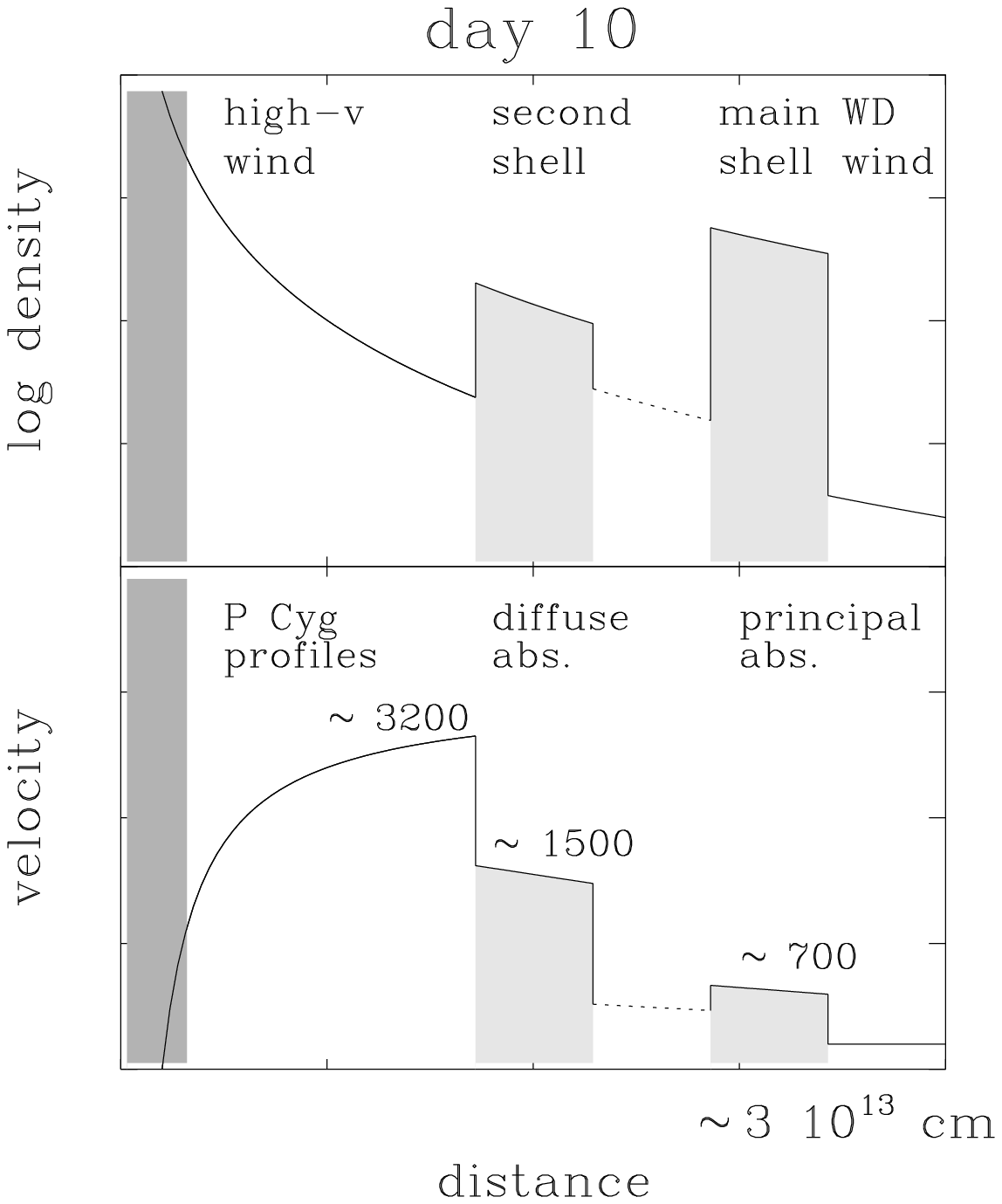,width=10cm}
\psfig{file=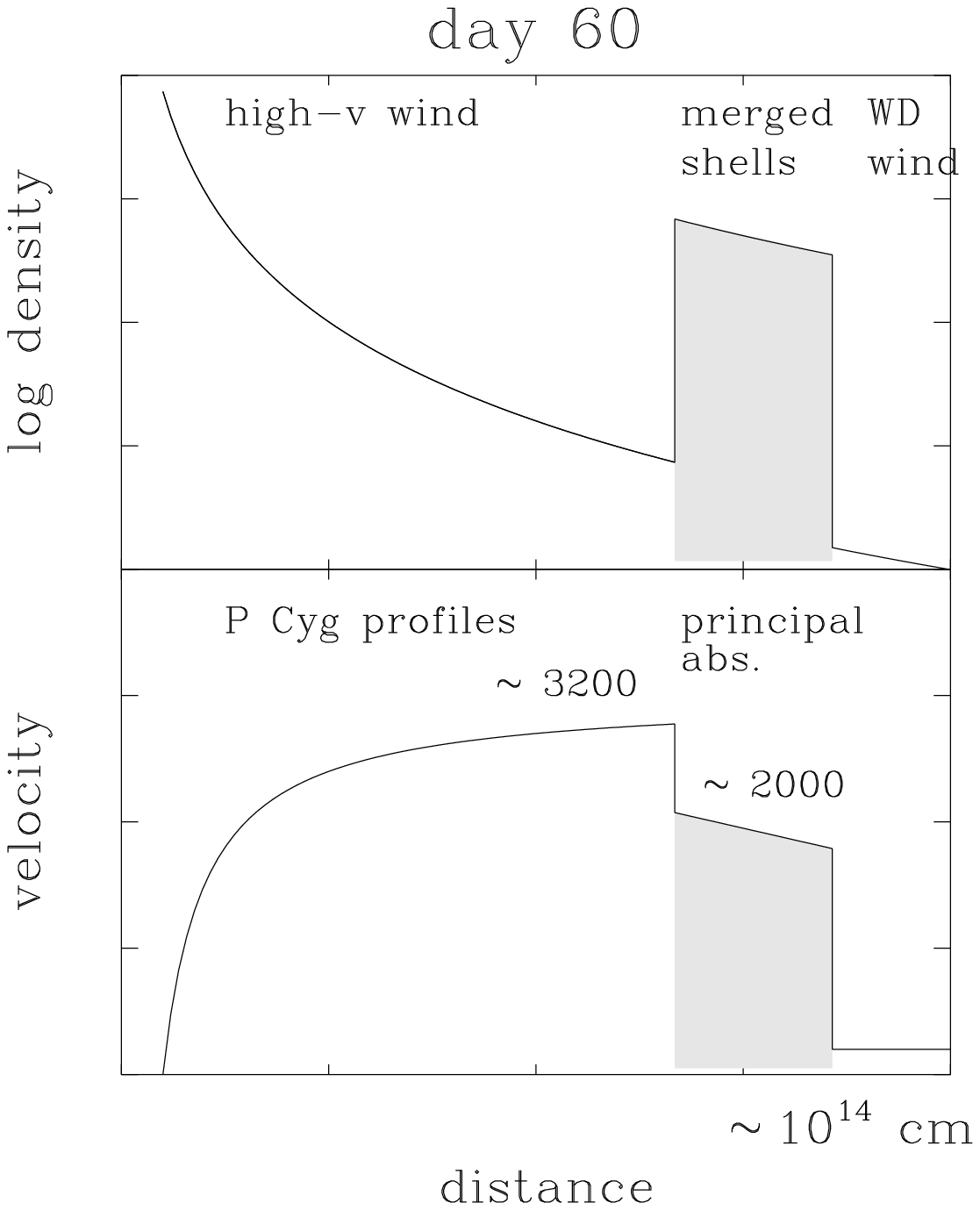,width=10cm}

\caption{A sketch of the density structure (top) and velocity
structure (bottom) of an empirical multi-phase wind model at about day
10, and day 60 described in Sect. \ref{sec:multi-phase}. The distance
scale is only approximate.  The different zones are indicated as well
as the approximate velocities.  On day 10 (left hand panel), the two
absorption line systems are formed in the two shells (light grey). In
this early phase the mass loss rate of the wind is so high that the
pseudophotosphere is located inside the wind. Radiation from the
deeper high density layers of the wind (dark grey) does not escape. On
day 60 (right hand panel) the high velocity wind is no longer
optically thick in the continuum, so the emission lines from the
deepest wind layers can be observed.  The second and the main shell
have merged into a single shell (light grey). The mass loss rate has
decreased so that the inner part of the wind is no longer optically
thick.}
\label{fig:modello} 
\end{figure*}

We stress that the velocity decreases outwards in both the main shell
and the second shell. This is {\it not} due to a deceleration of the
wind (which would be impossible, because the gravity of the white
dwarf is not strong enough to decelerate the wind once it has moved
away more than about 10 \Rsun\ or about $10^3~ R_{WD}$), but is due to
the fact that the ejection velocity increased with time, probably
in step with a decrease of the mass loss rate.

\subsection{The structure and evolution of the envelope}
\label{sec:time-evolution}

The evolution of the individual components of the nova envelope and
the observational consequences are better understood if the innermost
region, which provides the ionizing flux to the rest of the envelope,
is discussed first. Then, going from the inner to the outer regions,
the evolution can be described as follows:

%\begin{itemize}
\begin{description}
\item{ The {\it high velocity wind}} sets in after the other mass loss
phases.  It gives rise to the emission profiles that are seen as
prominent features after about day 6.  Up to roughly day 14 the mass
loss rate is so high that this wind is optically thick in the UV
continuum, so that the pseudo-photosphere lies within the wind.  As
the mass loss rate decreases, the wind becomes optically thin in the
continuum, and the lower layers of the pseudo photosphere move
inwards. This has two effects: (1) the effective temperature increases
as the effective radius of the star decreases, and (2) radiation from
the inner layers of the wind, where the velocity is smaller and the
density is higher than outside, becomes detectable.  This explains the
decreasing width of the emission components (Figs. \ref{fig:prof1},
\ref{fig:prof2} and \ref{fig:fwhm}), as well as the increase in the
degree of ionization in the wind with the higher ionization stages
peaking at later times than lower ionization stages
(Fig. \ref{fig:lowres}).

\item{\it The second ejection phase} at around day 3 to 5 produces the
{\it second shell} that has an initial velocity of about 1500 \kms\ or
less (Fig. \ref{fig:velocity}). The absorption lines of the diffuse
system are formed in this shell. During the ejection, which lasts a
short time, the ejection velocity slightly increases with time, which
results in an outward decreasing velocity. This explains why the
higher ionization absorption lines show a slightly more negative
Doppler velocity than e.g. the \ion{Fe}{ii} lines
(Fig. \ref{fig:stratifica}).  As the star gets hotter as a consequence
of the receding pseudo--photosphere, its increasing UV flux
accelerates the shell by radiation pressure in the very numerous UV
absorption lines, similar to the radiative acceleration of line--driven
winds.  This explains the increase in the radial velocity of the
diffuse system up to a velocity of $\approx$ - 2800 \kms\ near day 60
(Fig. \ref{fig:velocity}).  As the shell moves outward, its column
density decreases and the absorption lines get weaker
(Fig. \ref{fig:ferroew}) and narrower (Fig. \ref{fig:fwhm_fe2}).

At the same time, the degree of ionization increases because the star
gets hotter. This explains why the equivalent width of the Fe II lines
decreases, whereas the Fe III absorption lines first get stronger up
to day 40 and then get weaker (Fig. \ref{fig:ferroew}).
The decrease of the column density with time has another observable
effect. The region where the absorption lines are formed moves deeper
into the shell. This implies that the absorption occurs over a region
with a smaller velocity range. This explains the observed narrowing of
the absorption lines (Fig. \ref{fig:fwhm_fe2}).
 
\item {The {\it main shell}} is ejected at a velocity 
% of 700 \kms\ or less (Fig. \ref{fig:velocity}). 
of less than of 700 \kms\ (Fig. \ref{fig:velocity}) about 3 days before 
the discovery.
%
%\footnote{HENNY:  I changed 700 into  LESS THAN 700, and added
%when this happened; in other words, the ejection velocity at the 
%true outburst is LESS than that observed at day 0 }
%
The absorption lines of the principal system are formed in this
shell. There is a small negative velocity gradient, which explains the
difference in Doppler velocity of the absorption lines of different
ionization stages (Fig. \ref{fig:stratifica}). Radiation pressure
accelerates this shell in a similar way as the shell where the diffuse
absorption lines are formed, and the main shell reaches a velocity of
about - 1800 \kms\ near day 60. The degree of ionization slowly
changes with time (Fig. \ref{fig:ferroew}) in the same way as in the
second shell, and the absorption lines get narrower (see
Fig. \ref{fig:fwhm_fe2}) for the same reason.  The velocity difference
between this shell and the faster moving second shell inside leads to
a merging of the two shells around day 60, as explained in
Sect. \ref{sec:impulsive}.  This merging leads to the disappearance of
the diffuse system absorption lines after day 60, the jump in velocity
of the principal system absorption lines, of about 240 \kms, which 
occurs between day 60 and 80, and the detection of hard X-rays
around that time.  This situation at about day 60 is sketched in
Fig. \ref{fig:modello}.  The remaining shell will be overtaken by
material from the fast wind.  However, the density contrast between
the main shell and the fast wind may be so large that the dynamic
acceleration of the shell will take a long time. It will probably lead
to afragmentation, as is observed in old nova shells (Slavin, O'Brien
\& Dunlop 1995).

\item {The material ejected} by the white dwarf during the
pre-nova wind phase is little disturbed during the first few hundred
days. This material produces the very stable, weak and narrow
absorption lines seen at -600 \kms on top of the \ion{Mg}{ii} and the
\ion{N}{v} doublets (see Fig. \ref{fig:prof3} of Sect.
\ref{sec:n1240} and Fig. \ref{fig:prof1} of  Sect. \ref{sec:mg2}).
\end{description}
%\end{itemize}

\subsection{Interaction between the different wind phases?}
We notice that the overall outward decrease of the expansion velocity
in the different wind regions will lead to the wind layers catching up
with their neighbours.  Yet there is little evidence for merging of
the layers, except for the second shell merging with the main shell.
For instance, the fast wind with its velocity of about 3500 \kms\
observed on day 1.3, i.e. in between the two shell ejections, should
catch up with the main shell with its initial velocity of about 700
\kms, increasing to 1500 \kms, in about 10 to 15 days, if the main
shell was ejected 3 days before the fast wind started. The lack of a
clear sign of the interaction suggests a large density contrast
between the wind and the main shell.

One of the most significant results of this study is the simultaneous
increase in velocity of the main shell and the second shell.  We have
explained this in terms of radiative acceleration by the star due to
its increasing UV flux. Could the acceleration be due to dynamic
interaction (momentum transfer) of the different shells?  This seems
unlikely, given the fact that the acceleration of both shells up to
day 60 is remarkably similar (Fig. \ref{fig:velocity}). If the main
shell was accelerated by the interaction with the second shell, and
the second shell by interaction with the fast wind, the similar
acceleration would require a finetuning between the momentum increase
of the two shells by the two interaction zones, which seems
unlikely. On the other hand, if the acceleration of the two shells is
due to radiation pressure, the similar acceleration is simply the
result of radiation pressure by optically thin lines in both
shells. For radiation pressure in optically thin lines the radiative
force per unit mass, and thus the acceleration, is proportional to the
flux and independent of the density.  So shells with different
densities, but with rather similar ionization conditions, will
experience the same acceleration. Still, the shell closer to the star
will experience a slightly larger acceleration because of the larger
incoming flux. The different radial distance of the shells explains
why the second shell, which is closer to the star than the main shell,
experiences a slightly larger acceleration.

An interesting question is how the above interaction would affect the
velocity field of the pre--maximum matter. The only trace we have of
the pre--nova wind is the presence of narrow and stable components of
the \ion{N}{v} doublet at -600 \kms appearing superimposed on the
doublet emission. The stability in velocity and equivalent width of
these narrow components between day 123 and 192 is not suggestive of
formation by dynamical interaction, but rather supports the idea that
they are formed by photoionization in a low density region far away
from the nova.

%**********************************************************************
%**********************************************************************

\section{Discussion and conclusions}
\label{sec:discussion}

{The present investigation, based on high resolution ultraviolet
spectroscopy of V1974 Cyg has led  to an empirical model in which
the object goes through different mass loss phases. The major
results of this study are:}\\

(1) In the earliest days the  high velocity wind is optically
thick so that the pseudo--photosphere lies within the wind itself. The P
Cygni profiles observed in the strong UV resonance lines are formed in
this wind. The absorption components of the P Cygni profiles do not
start at zero Doppler velocity, because the lower velocity or
acceleration layers of the wind are located below the
pseudo-photosphere. As the mass loss rate of the wind decreases, the
pseudo-photosphere moves inwards, and the deeper higher density wind
layers become visible, where emission components are formed by
recombination and/or collisional excitation.  The receding
pseudo-photosphere also results in a decrease of the effective radius,
which causes an increase of the effective temperature.  This explains
the increasing degree of ionization, reflected in the observed
strengthening of the higher excitation emission lines at the same time
that the lower excitation lines become fainter. As the mass loss rate
and the wind density decrease further, the emission comes from deeper
layers in the wind, where the velocity is smaller, which results in
the observed narrowing of the emission lines (see Williams 1992).

(2) Superimposed on the emission lines, one observes two
absorption line systems, which are formed in two different shells.
The main shell is ejected first and the second shell is ejected later,
around day 3 to 5 (see spectra of Andrillat \& Houziaux, 1993). 
%%%%%%%%%%
% INTERESTING 
%The ``hickup''
%in the lightcurve at day 7 (Annuk et al. 1993) might mark the end of
%this ejection phase.
The acceleration of the two shells is remarkably similar, which
suggests a common mechanism.  We propose that the shells are
accelerated by radiation pressure in spectral lines (similar to the
winds of hot stars) due to the increasing UV brightness of the
star. 

(3) Within each of the two shells the higher ionization lines have a
slightly higher outflow velocity than lower ionization lines, which 
indicates  that the velocity of the shells  decreases slightly
outwards. This is not due to the gravity of the star, nor to the
interaction with material outside the shells, but most likely due to
the fact that the velocity slightly increased with time during the
ejection of the shells.

(4) The inner shell (second shell), which has a higher velocity than
the outer shell (main shell) catches up with the outer shell at about
day 50 to 70. This agrees roughly with the interaction time expected,
based on the difference in velocity of the two shells
(Sect. \ref{sec:impulsive}).  Also, it explains the rather drastic
disappearance of the faster shell system around that date, as well as
the sudden increase in velocity of the main shell, and the appearance
of hard X-ray flux. From the velocities of the faster secondary shell,
of the slower main shell, and the final velocity of the merged shell
we estimate that the mass ratio of the main shell to the second
shell is  roughly 4/1 (Sect. \ref{sec:impulsive}).

(5) In Sect. \ref{sec:interpretation} we estimated the distance to
V1974 Cyg by assuming a spherically symmetric expansion for the nova
envelope at the time of the H${\alpha}$ interferometric measurements
of Quirrenbach et al. (1993) on day 10, and by combining these with
the estimated radius of the principal system at the same date. The
value so obtained, 2.9 kpc, is in good agreement with the independent
estimates by Quirrenbach et al. (1993), Della Valle \& Livio
(1995) and Paresce et al. (1995).\\

Our model is qualitatively in agreement with the optically thick
wind models of novae, by Kato \& Hachisu (1994), first introduced by
Friedjung (1966a,b,c), and further treated by Kovetz (1998) and Bath
\& Harkness (1989), who showed that the decreasing mass loss rate of
the wind will result in an inward moving photosphere, an increase in
the effective temperature and an increase in the degree of ionization
in the wind.  Since in these models matter is accelerated deep inside
the pseudo--photosphere, it is possible in this way to explain the
formation of the peculiar P Cygni profiles seen in the earliest
spectra and the progressive narrowing of the emission lines.  Also,
because of the predicted inward recession of the pseudo--photosphere
and the consequent increase of the effective temperature, it is
possible to explain with an unique mechanism the acceleration of the
two outward moving shells by radiation pressure in the UV lines.  The
stratified structure within the expanding shells
%
%\footnote{HENNY: I had to change this because this does not 
%apply to the inner fast wind}
% 
as suggested in the present model, is in agreement with the major
observational features reported by Friedjung (1966b,c), i.e. the
overall decrease of the expansion velocity and of the (Zanstra)
temperature with increasing radial distance.
       
%%%%%%%%%%% 
%The present model is also 
%consistent with the major observational features reported by Friedjung
%(1966b,c), i.e. the overall decrease of the expansion velocity and of
%the (Zanstra) temperature with increasing radial distance.}

However, our observations and our empirical model require that
the nova ejected two shells: one before the discovery at about day -3,
and one at about day 3 to 5.  In this way it is possible to explain
why the absorption of the principal system lines is superimposed on
the P Cygni profiles of the wind observed on day 1.3 and why the
principal and diffuse--enhanced system lines are superimposed on the
wind emission lines after day 6.  The first shell contains most of
the ejected matter and the second shell, ejected with a higher
velocity, contains considerably less material (about 1/4 in this
case).  We might suggest that the second ejection is the reaction to
the main ejection. After the ejection of the main shell, the envelope
is out of balance and will try to restore equilibrium, but if
this does not succeed immediately, due to some sort of overshooting, a
second, less massive shell might be ejected.  We like to point out
that a similar effect occurs in the large eruptions of luminous blue
variables (LBV), which also show that the main eruption is followed by a
second one (Humphreys, Davidson  \& Smith, 1999).  In the case of
LBVs, which are very extended and have a small surface gravity, the
main eruption lasts for tens of years, and the second eruption occurs
about 60 years after the main one. In the case of eruptions from the
surface of white dwarfs, where the gravity is very high, the time
scales will be much shorter. It might be interesting to see if
dynamical models of nova eruptions show this effect.

According to Friedjung (1987), the principal system shell (here the
main shell) is formed by interaction of a wind continuously ejected at
a decreasing rate, with material ejected before optical maximum. This
model does not explain, however, the presence of the second absorption
system with higher velocity, produced inside the main shell. Although
Friedjung's interaction model is attractive because wind--shell
interaction is likely to occur (see also McLaughlin, 1960; Friedjung
\& Duerbeck 1993), it is not clear how this could lead to {\it two}
shells with strongly different and well separated velocities. On the
other hand, as discussed in Sect.  \ref{sec:impulsive}, it is unlikely
that the two absorption systems are due to flows in two directions,
e.g. one in the orbital plane of the binary and one in the polar
direction.

To conclude, we would like to encourage further studies of radial
velocity variations in novae  (poorly pursued nowdays), especially
during the first few days, since they
represent a powerful means to get insight into the dynamical structure of
the nova envelopes and into the origin of the X--ray emission. Such
studies should also provide valuable input information for non--LTE
synthetic spectra of novae (see e.g. Short et al. 2001).

\begin{acknowledgements}
We are very grateful to Dr. Roberto Viotti for his numerous and
helpful comments and to Dr. Michael Friedjung for stimulating this
investigation and for constructive criticism. Finally, we are grateful
to the anonymous referee for several useful comments and suggestions.

\end{acknowledgements}

\end{document}